# Is normalized biomass really abundance? pitfalls and misconceptions in the field of size spectra analysis: a case for back-transformed spectra and standardized binning

**Ralf Schwamborn**

Oceanography Department, Federal University of Pernambuco (UFPE), Recife, Brazil

*e-mail*: ralf.schwamborn@ufpe.br

## Highlights

1. The widely used NBSS method has non-biomass units, leading to ambiguity.
2. The new, robust backtransformed bNBS method uses convenient, intuitive biomass units.
3. Simulations show NBSS and bNBS both scale perfectly as biomass (negligible bias).
4. Normalization methods, biomass density functions, and carrying capacity are discussed.
5. Standardized binning and units are urgently needed for ecosystem size spectra research.

# Graphical Abstract

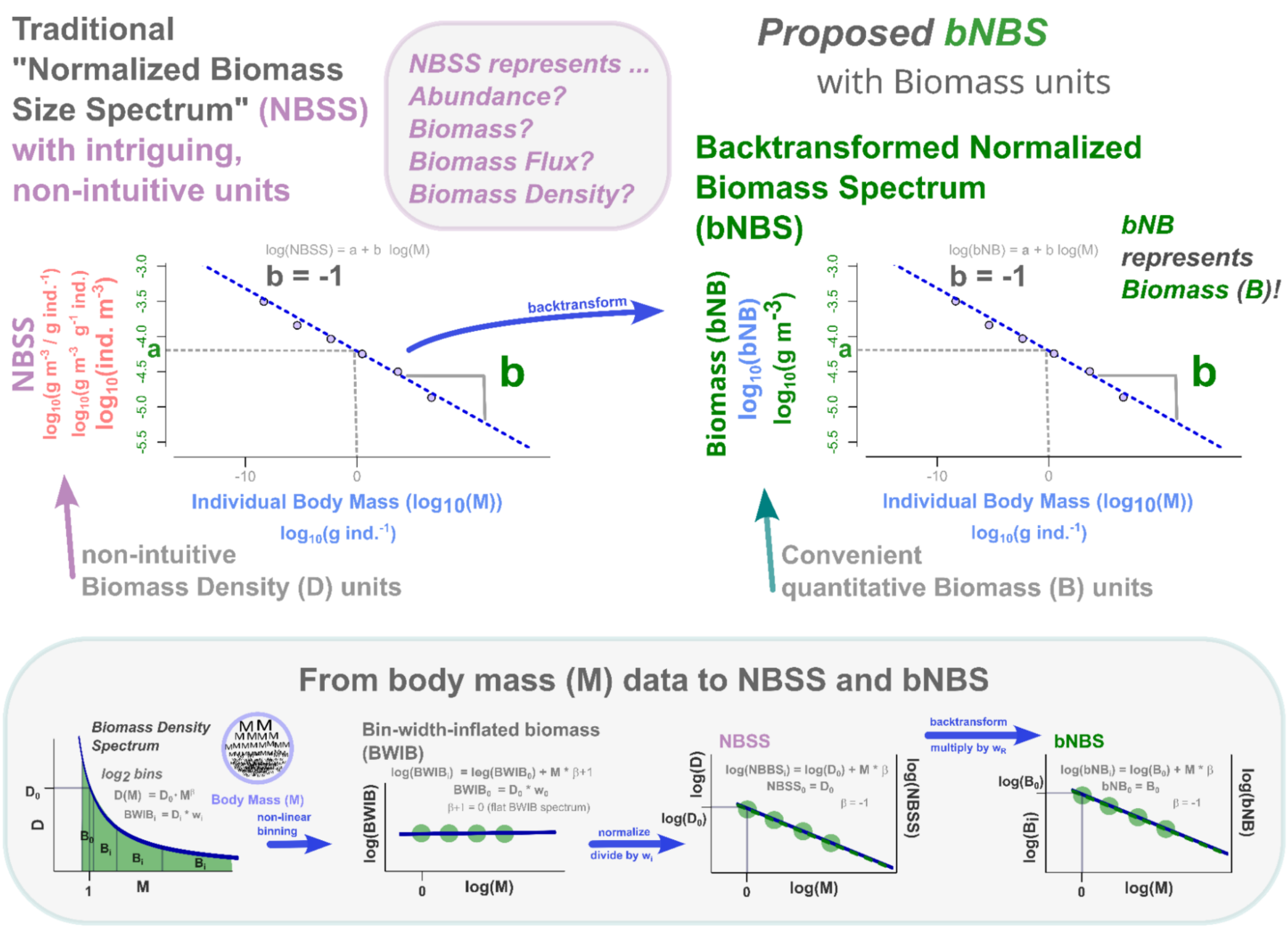

# Abstract

The NBSS (normalized biomass size spectrum) is a common, intuitive approach for the study of natural ecosystems. However, very few studies have been dedicated to verifying possible flaws and paradoxes in this widely used method. Evident points of concern of the NBSS method are 1.) the loss of variability due to binning and 2.) the use of intriguing non-biomass units (i.e. abundance units). The main objectives of this study were to verify, test and analyze the procedures involved in transformations that lead to the NBSS plot, and to check for the correctness of currently used units, while testing the hypothesis that NBSS indeed represents biomass density, not abundance, while developing i.) a new conceptual framework, ii.) a new terminology within "Size-spectrum Inference, Scaling, and Estimation" (SIZE) theory, iii.) a novel back-transformation method, iv.) high-resolution kernel density estimation (KDE) plots of the density function shape, and v.) a new calculation method for numerical values, dimensions, and units. Extensive tests with *in situ* and synthetic (simulated) data were used to compare the original biomass distributions with binned outputs. Original biomass units (g $m^{-3}$) and dimensions are retained in the proposed robust, bootstrapped "backtransformed and normalized biomass spectrum" (bNBS). Previously published common NBSS may be simply converted to quantitative biomass B by multiplying with a reference unity bin with (wR = 1 mass or volume unit). If a standardized framework (standard mass and space units, standard reference bin width $w_R$) is used, the proposed bNBS may allow for a new approach of robust size spectra science, that allows for quantitative inter-comparisons of abundance and biomass across size ranges, regions, and time periods.

## 1. Introduction

Obtaining quantitative estimates of abundance and biomass in natural habitats is key to understanding our planet, and the base of all ecosystem models. Yet, this task is far from trivial. Additionally to practical challenges (e.g., sampling organisms that are difficult to detect or capture), one fundamental and intriguing problem in any *in situ* quantitative sampling effort is that biomass and abundance estimates are a function of the arbitrary choices of target size and size range considered. Targeting smaller organisms (or adopting broader size ranges) will generally result in higher estimated biomass and abundance values.

Thus, the analysis of size spectra (i.e., the relationship between body size and abundance, Sheldon et al., 1972, 1973, 1977) and the elaboration of size spectra-based ecosystem models (e.g., Dalaut et al., 2025) have emerged as central topics in quantitative ecosystem research. Marine (e.g., Sheldon et al., 1972), freshwater (e.g., Rossberg et al., 2019), and terrestrial ecosystems (e.g., Mulder et al., 2008) generally present a power-law-shaped size spectrum characterized by vast numbers of small organisms (e.g., bacteria), while the largest ones (e.g., blue whales) are extremely rare. Thus, it is impossible to visualize and analyze natural size distributions with a histogram of regularly spaced size bins. Such a regularly spaced histogram would inevitably lead to many empty bins (gaps) within the rare, large size classes (Suppl.

Mat., Fig. SM1a). Empty bins are indeed a nuisance in histogram-based visualization and analysis, with a considerable effort being dedicated to finding the optimal bin widths (Freedman and Diaconis, 1981, Wand, 1997, Silverman, 2018, Schwamborn et al., 2025). Therefore, the use of a binning vector with non-linear (geometrically increasing) bin widths is necessary to adequately represent power-law distributions in a histogram, given the rarity of large-sized organisms.

The most common method to describe a size spectrum by non-linear binning, that has been applied in an immense wealth of publications (see reviews, e.g. in Sprules & Barth, 2016, Atkinson et al., 2024, Ersoy et al., 2025) is the NBSS ("*normalized biomass size spectrum*", Platt and Denman, 1977, Fig. 1). NBSS are widely used for the description of global ecosystems and for the validation of modern individual-based ecosystem models (e.g., Dalaut et al., 2025). Hereby, "*normalized*" means that the binned biomass values are divided by the bin width (i.e., the body mass range) of each histogram bin.

The main advantage of the NBSS method is that it allows for the use of non-linear binning (with wider bins at larger sizes), enabling an adequate representation of large-sized organisms. Another, often overlooked advantage of the NBSS method is that log-linear slope and intercept are invariant to the binning vectors used, due to the normalization by bin widths. Conversely, for all other, non-normalized binning methods, numerical "y" values (and hence the intercept of linear models) are a function of bin width.

Most importantly, non-linear binning has a well-known distorting effect on the represented biomass: as bin width (w) increases for larger organisms, it artificially inflates the estimated biomass of those large organisms. The resulting distorted representation of the biomass-body mass relationship may be called the "bin-width-inflated biomass" spectrum (BWIB, Fig. 2). To correct this distorting effect, during the calculation of the NBSS, each $BWIB_i$ value in each bin "i" is divided by its corresponding bin width $w_i$:

$$NBSS_i = BWIB_i / w_i$$

Simply put, the NBSS is a popular way of correcting for the distortion due to non-linear binning (Fig. 2).

Power-law-shaped ecosystem size spectra are easily recognized by a linearly declining shape, when plotted on a double logarithmic scale (Fig. 1). In the (often observed) case of a power-law spectrum with a log-linear NBSS slope of $b = -1$, the use of geometrically increasing bin sizes will lead to a perfectly flat (slope = 0, which is not a power-law distribution), unrealistic BWIB *vs* body mass spectrum that has been used by several authors to visualize a supposedly "*flat size spectrum*" in natural ecosystems (e.g., Boudreau and Dickie, 1992, Brown et al., 2004, Maxwell & Jennings, 2006, Trebilco et al., 2015, Fock et al., 2026). This widespread error has led to the widely cited erroneous assertion that biomass is supposedly "*evenly distributed*" (e.g., Boudreau & Dickie, 1992; Maxwell & Jennings, 2006; Trebilco et al., 2015) or "scaled to $M^0$" (Brown et al., 2004; Fock et al., 2026) across size classes in aquatic ecosystems.

Conversely to the supposedly "*flat*" BWIB, the NBSS produces a conveniently binned power-law distribution, with a linearly downtrending slope of approximately $b = -1$ on a log-log-scaled plot. Based on the immense wealth of NBSS data and literature, the generally prevailing view in current scientific literature (see e.g., Dugenne et al., 2024, Schwamborn, 2025) asserts that the NBSS slope of b = -1 represents the correct and accurate *in situ* biomass-body-mass scaling of natural ecosystems (Fig. 1).

Yet, few studies have been dedicated to verifying possible bias, flaws, and paradoxes in the widely used NBSS method. An evident shortcoming of this method is that it is not based on the original data, but rather on grouped (binned) data, where there is possibly an effect of the binning process on the outcome of the analysis. The most obvious effect of binning is the loss of variability, which evidently should raise concerns about whether the most popular currently used representation method for size spectra in marine science (the NBSS), faithfully captures the true shape of the original data - such as its peaks, bumps, and troughs.

Edwards et al. (2017, 2020) highlighted possible binning artifacts, and compared binning-based methods with the maximum likelihood estimation (MLE) of the size spectrum slope, without binning. In their simulations, MLE proved superior to several binning-based methods, which had considerable bias in slope and variability estimates. The strong claims made in both studies (Edwards et al., 2017, Edwards et al., 2020) about the superiority of the MLE method over binning-based approaches have led to confusion and uncertainty about the validity and correctness of NBSS (a binned method), even though NBSS was neither applied nor tested by Edwards et al. (2017, 2020).

However, one important drawback of their MLE method is that, while a being a useful estimator of the slope (b), it is not capable of determining the intercept (a) of the spectrum, which highlights the fundamental challenge of quantifying absolute levels of biomass and abundance in any power-law structured natural ecosystem.

---

**Normalized Biomass Size Spectrum (NBSS)**

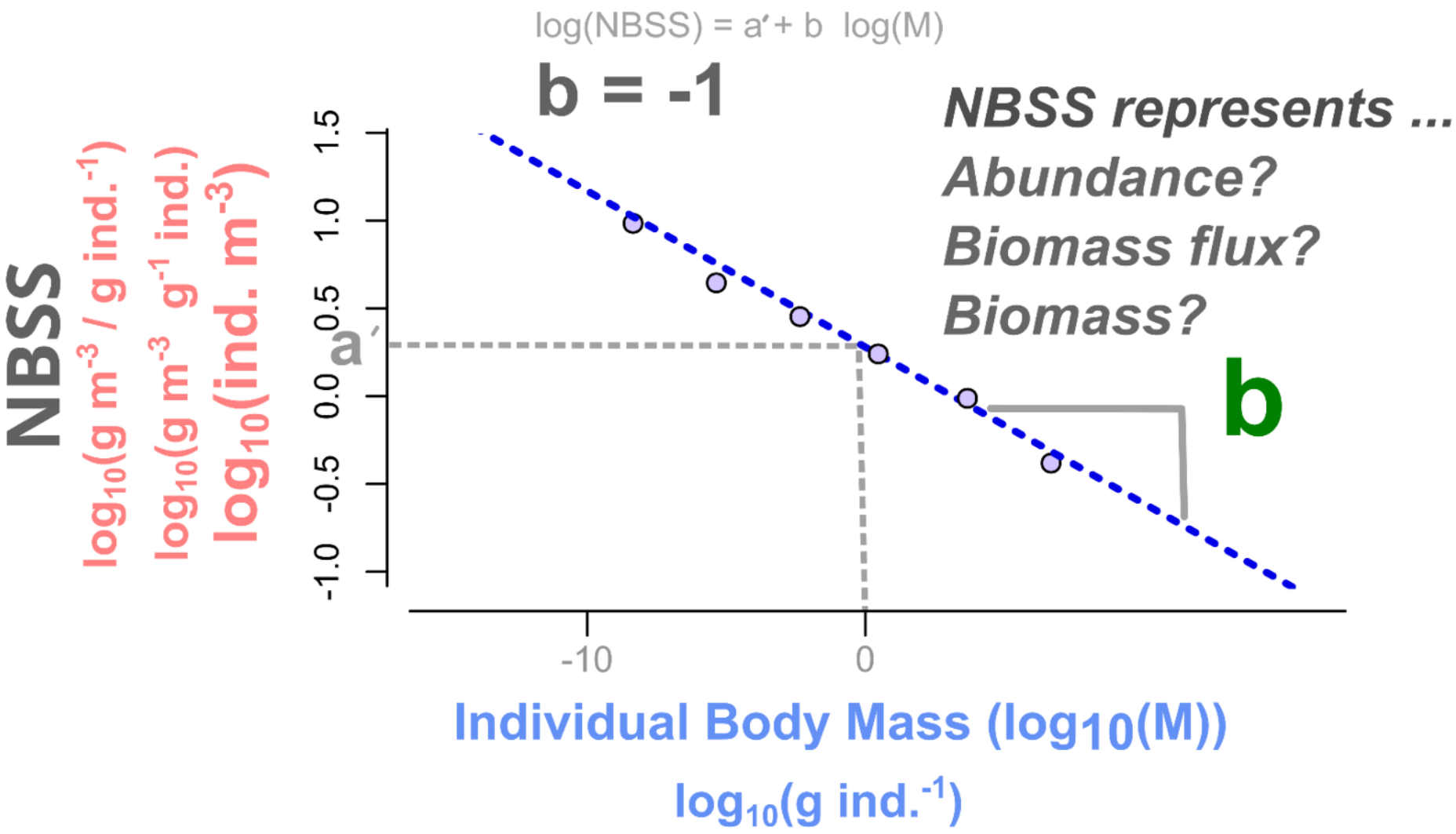

**Figure 1:** Currently used NBSS (normalized biomass size spectrum) method. Note the abundance (or "biomass flux") units on the y axis.

The most striking paradox of the NBSS method, is that the units of the plot are actually units of abundance, not biomass (Fig. 1). When dividing the original biomass data (e.g., g $m^{-3}$) by the bin width of each body mass bin (e.g., g $ind.^{-1}$), the resulting NBSS data appear to be in units of abundance (Fig. 1). Many authors (e.g., Platt and Denman, 1978) have thus concluded that the NBSS actually represents the shape of the *Abundance (!) vs* body mass relationship, not a biomass spectrum.

For instance, the units of a common NBSS are: (g / $m^3$ ) / (g / indiv.) = g $m^{-3}$ $g^{-1}$ indiv., which simplifies into NBSS = indiv. / $m^3$ (!).

Such exotic "abundance" or "pseudo-abundance" units can be observed in virtually all NBSS publications on biomass (!) spectra , on the y-axis of all plots (Fig. 1). Far from being a merely formalistic issue, the "NBSS-Biomass-Abundance-paradox" (NBSS are biomass data, but have abundance units) has far-reaching consequences for our understanding of our planet. For instance, only biomass (not abundance) represents the mass and energy stored in living organisms and reflects their contribution to food webs, energy flow, ecosystem productivity, and trophic structure. Also, abundance and biomass spectra exhibit fundamentally different scaling relationships with body mass. In pelagic marine ecosystems, a slope of approx. b = -2 can be expected for abundance-body mass spectra (Figueiredo et al., 2025), rather than approx. b = -1, as observed for most pelagic biomass-body mass spectra (Dugenne et al., 2024).

Alternatively, some authors (e.g., Platt and Denman, 1978, Thompson et al., 2013) have looked the NBSS, with its awkward units, not as representing biomass or abundance, but as a representation of "biomass flux", i.e., the rate of change of total biomass with individual mass (dB/dM). Under this perspective, the common NBSS could be regarded as a biomass-change-body-mass spectrum (dB/dM vs M). Since the first days of size spectra science (Platt & Denman, 1978) several authors have interpreted the NBSS values as a measure of "biomass flux", i.e., rate of biomass flow through the food web (e.g., Platt and Denman, 1978, Thompson et al., 2013), or as an often ill-defined form of “biomass density” (e.g., Boudreau and Dickie, 1992; Benoit and Rochet, 2004; Cuesta et al., 2018).

Analogously, the NNSS (normalized numbers size spectrum, Blanco et al., 1994, Vandromme et al., 2012, Figueiredo et al., 2025, Fig. 3), a popular form of abundance *vs* body mass-spectrum, does actually not have abundance units (e.g., ind. $m^{-3}$), but rather awkward "abundance flux" or “abundance density" units (abundance per body mass, e.g., “(ind. / $m^3$ ) / (g / ind.) = ind. $m^{-3}$ $g^{-1}$ ind., or ind.$^2$ $m^{-3}$ $g^{-1}$).

An evident statistical flaw of the currently used NBSS approach, that has hitherto been ignored, is that the degrees of freedom from original numbers are lost in the binning process. The linear regression treats the binned data without accounting for the original sample sizes (for instance, a linear regression on a NBSS histogram based on 100 individuals produces exactly the same confidence intervals as one based on thousands of individuals, an evident statistical absurdity).

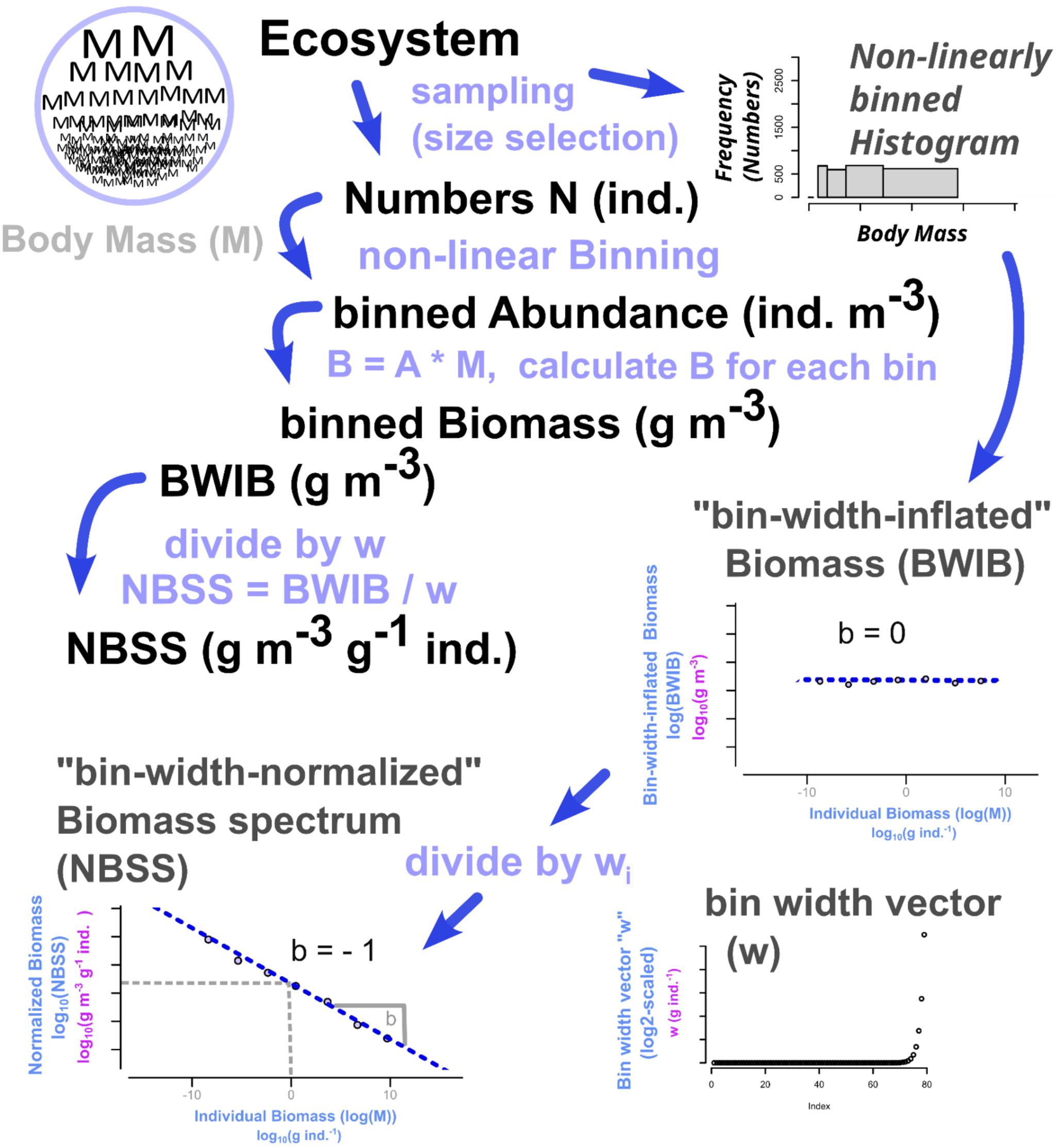

**Figure 2:** Calculation procedure for the common NBSS (normalized biomass size spectrum) method. BWIB (bin-width inflated biomass), an intermediary calculation step towards the NBSS, is often indirectly obtained through non–linearly binned, bin-width inflated Abundance BWIA, which is multiplied by midpoint body mass M, where BWIB = BWIA * M. NBSS is then obtained from BWIB through dividing BWIB by the non-linear bin width vector "w" , where NBSS = BWIB / w.

Also, the linear model fit algorithms for NBSS usually apply common ordinary least squares regression (OLSR), an outdated method that is well known to be prone to non-linear outlier effects, that become stronger when outliers are at the lower or upper extremes of the dataset. Alternatively, there are many modern fit methods, such as robust regression (e.g., Huber, 1964; Rousseeuw and Leroy, 2003), that are specifically designed to limit outlier influence, but are rarely used in NBSS-based studies (Schwamborn et al., submitted). Still, common NBSS studies treat histogram bins as if they were samples, and do not consider the uncertainty from the original binning procedure. A statistically robust multi-step bootstrapping routine (as in Schwamborn et al., 2019) that accounts for the original sample size (and all sources of uncertainty in the steps leading to the NBSS linear regression slope "b"), proposed herein, is necessary for the calculation of proper confidence intervals.

Another hitherto ignored, potentially severe problem of the NBSS method is that the universal log-log-linear slope of b = - 1 may be an artifact of model construction and linearizing artifacts (Schwamborn 2018). This would explain the perfectly linear shape of many size spectra (e.g., Dugenne et al., 2024) and the ubiquity of the startlingly precise b = 1 value for NBSS slopes (e.g., Dugenne et al., 2024, Ersoy et al., 2025). Most surprising, the extensive Dugenne et al. (2024) plankton NBSS study also may suggest that there is a universally constant intercept (i.e., biomass) in pelagic ecosystems across the globe (including tropical, temperate, and polar regions, that are obviously very different in plankton biomass), which additionally raises questions about possible artifacts and issues in the analysis of outputs of NBSS linear models.

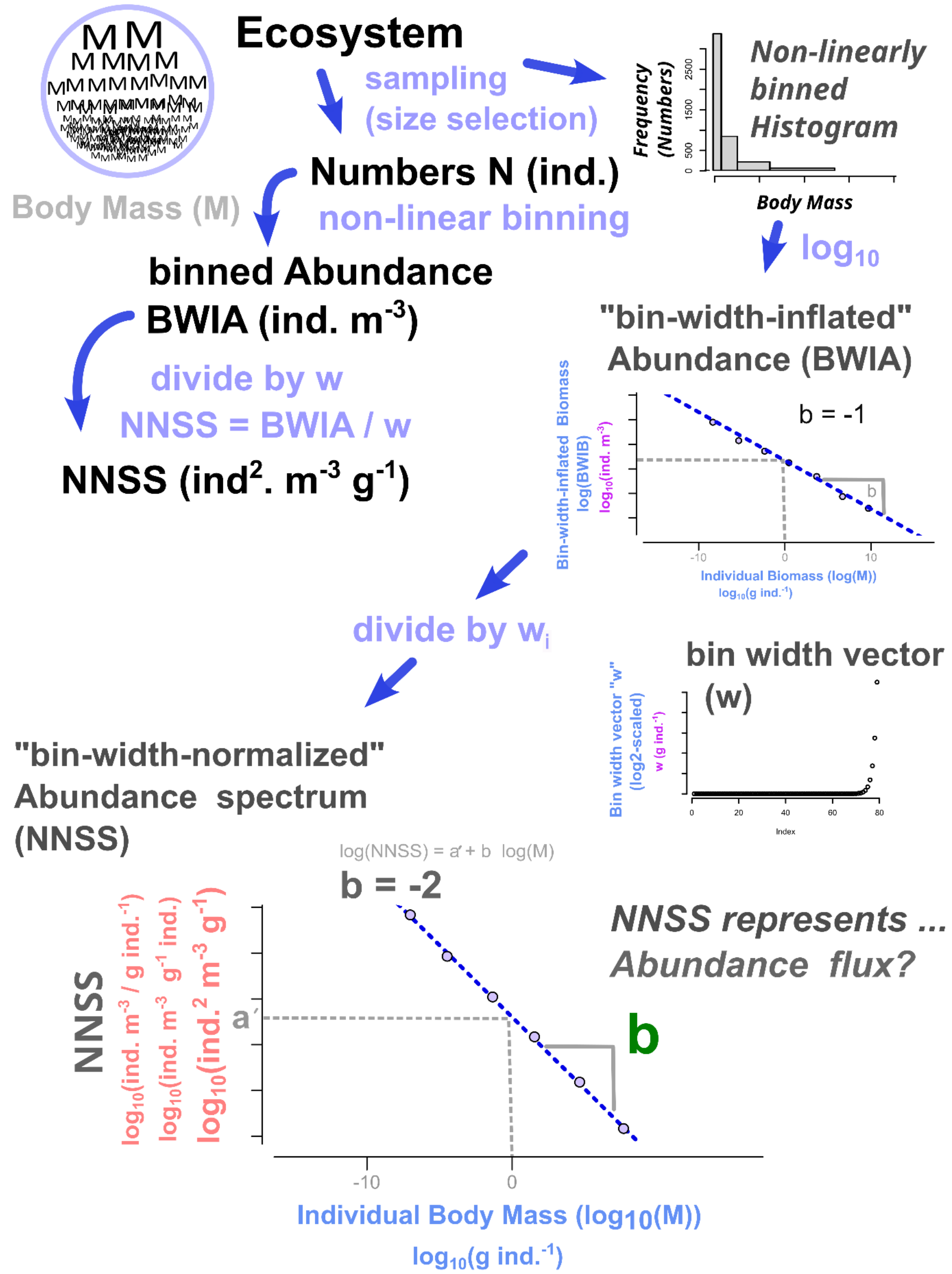

**Figure 3:** The path from body mass (M) data to the NNSS (normalized numbers size spectrum), a form of “abundance *vs* body mass spectrum”. Note that the “y” axis does not have common abundance units (e.g. ind. $m^{-3}$), but complex "abundance flux” units (abundance per body mass or volume, e.g. , ind. $m^{-3}$ / g $ind.^{-1}$ <=> $ind.^2$ $m^{-3}$ $g^{-1}$).

A possible issue of the NBSS approach, that may lead to severe bias, is the use of non-independent variables in model construction, known as spurious autocorrelation artifacts (SAA, Pearson, 1897). Numerous studies have shown that SAA represents a well-described, pervasive, and deleterious issue in quantitative science (Pearson, 1897, Reed, 1921, Chayes, 1949, Bensen, 1965, Kenney, 1982, Kanaroglou, 1996, Brett, 2004, Auerswald et al., 2010, Schwamborn, 2018). As a consequence, many popular calculation methods have faced severe criticism, e.g., for linearizing transformations that always produce apparently perfectly linear plots (even for bogus data), with erroneous results and biased models (Schwamborn, 2018). About four decades ago, Prothero (1986) presented a simulation study that contained a supposed proof that the NBSS slope of b =-1 may be a spurious artifact, but these simulations have been heavily criticized by a subsequent study (Blanco et al., 1994). Since then, only few validations and in depth verifications of the principles and possible bias in the NBSS method have been conducted. For instance, Gómez-Canchong et al. (2013) investigated the strong relationship between NBSS slope and intercept within time series, a classical case of SAA.

In many cases, standard methods have been ultimately discarded due to severe statistical artifacts, such as SAA (Schwamborn, 2018) and had ultimately to be replaced with SAA-free methods (Schwamborn et al., 2019). In NBSS analysis, total Biomass “B” in each size bin may be indirectly reconstructed from Abundance A and individual mass M (as B = A * M), after which log(B) is plotted against log(M) for linear model fitting. Thus, B and M are possibly not independent, potentially producing a form of linearizing SAA bias. Another potential issue is the normalization procedure, where each biomass B value is divided by bin width w.

Since both data sets, B and w, contain units of mass, there is the possibility of SAA due to the calculation of NBSS = B / w. The use of a y = 1/x model, where y is calculated from x, produces an inversely proportional relationship, that is identical to a power-law distribution with a slope of b = -1, generally regarded to be a universal NBSS slope value in pelagic ecosystems (Dugenne et al., 2024, Schwamborn 2025). Thus, extensive simulations were conducted in this study to verify for bias and possible spurious effects and linearizing artifacts.

The blatant "NBSS-Biomass-Abundance-paradox" has not yet been explicitly analyzed by previous studies. Nevertheless, it has led to a lot of confusion regarding the interpretation of NBSS plots and and to the development of what has been referred to as an "NBSS theory" (Platt & Denman, 1978, Blanco et al., 1994, Marcolin, 2013, Marcolin, et al., 2013, Hernández-Moresino et al., 2017), reflecting the inherent complexity of this subject. In the context of "NBSS theory", an often stated misconception is that the NBSS linear model intercept should be a representation of primary production, or "ecosystem productivity", and NBSS should represent biomass flux (e.g., Zhou, 2006, Marcolin et al. 2013, Sato et al., 2015).

As a simple, intuitive analogy, distance (a one-dimensional variable) is very different from speed (speed = distance / time). Following the line of thought of "NBSS theory", the NBSS would represent biomass flux, which is analogous speed (the transit rate of mass along the spectrum), and biomass B would not be adequately represented in the NBSS plot. This rationale may be most precisely summarized as the "NBSS-is-biomass-flux" hypothesis.

As an alternative hypothesis, one may argue that i.) the distortion due to non-linear binning and ii.) the subsequent bias-correction (dividing by w to obtain the NBSS) both just represent simple forms of bias-insertion and bias-correction. Under this alternative, "NBSS-is-biomass" hypothesis, NBSS actually represents biomass (not the change of biomass with individual mass). If the "NBSS-is-biomass hypothesis" is correct, then the units, dimensions (and numerical values), of all previously published and currently utilized NBSS plots must urgently be corrected, but the shape of the plots and the log-log-linear slope of the NBSS (approximately b = -1) are probably correct.

Somehow implicitly acknowledging this paradox, most published NBSS datasets, including the most recent extensive NBSS studies (e.g., Dugenne et al., 2024, Fock et al., 2026, Schwamborn et al., submitted) present and interpret the NBSS as a direct representation of biomass B, in spite of exotic, non-biomass units (e.g., units of abundance, or biomass flux). The "NBSS-Biomass-Abundance-paradox" leads us to pose the question: Does NBSS represent biomass, abundance, or biomass flux?

Further contributing to ambiguity and confusion, the current terminology is clearly inadequate for the study of biomass-body-mass relationships and spectra. First, the very name NBSS contains the wording "size", which implies a relationship between body size (not body mass) and some other variable. Conversely, common NBSS plots show body mass (or body volume), not body size, on the x-axis. Secondly, and most seriously, all existing NBSS studies deem synonymous 1.) the name of the size spectrum plot (i.e., the NBSS is the "NBSS vs M" relationship) and 2.) the name of the normalized biomass variable itself (NBSS is the name of "y" values of the NBSS plot). All reviewed papers use the same term (NBSS) for the plot, the log-linear model, and the normalized variable itself. This unusual conflation may have contributed to the ongoing confusion, ambiguity and uncertainty over whether NBSS refers to biomass, to a biomass index, or to "biomass flux" (dB/dM).

The main objectives of this study were to verify, test and analyze the procedures involved in transformations that lead to the popular NBSS plot, and to check for the correctness of currently used units, while testing the hypothesis that NBSS indeed represents biomass (i.e., the "NBSS-is-biomass" hypothesis), not abundance or biomass flux (dB/dM). The main challenges herein were i.) to develop a new conceptual framework, ii.) new terminology, iii.) a novel back-transformation method (i.e., to find the optimal back-transformation factor or vector), iv.) a simple, new calculation method (the novel bNBS plot and model, Fig. 4), that yields the best (i.e., least biased) representation of the original biomass vs body mass distribution shape, numerical values, uncertainties (95% confidence intervals and envelopes), dimensions, and units.

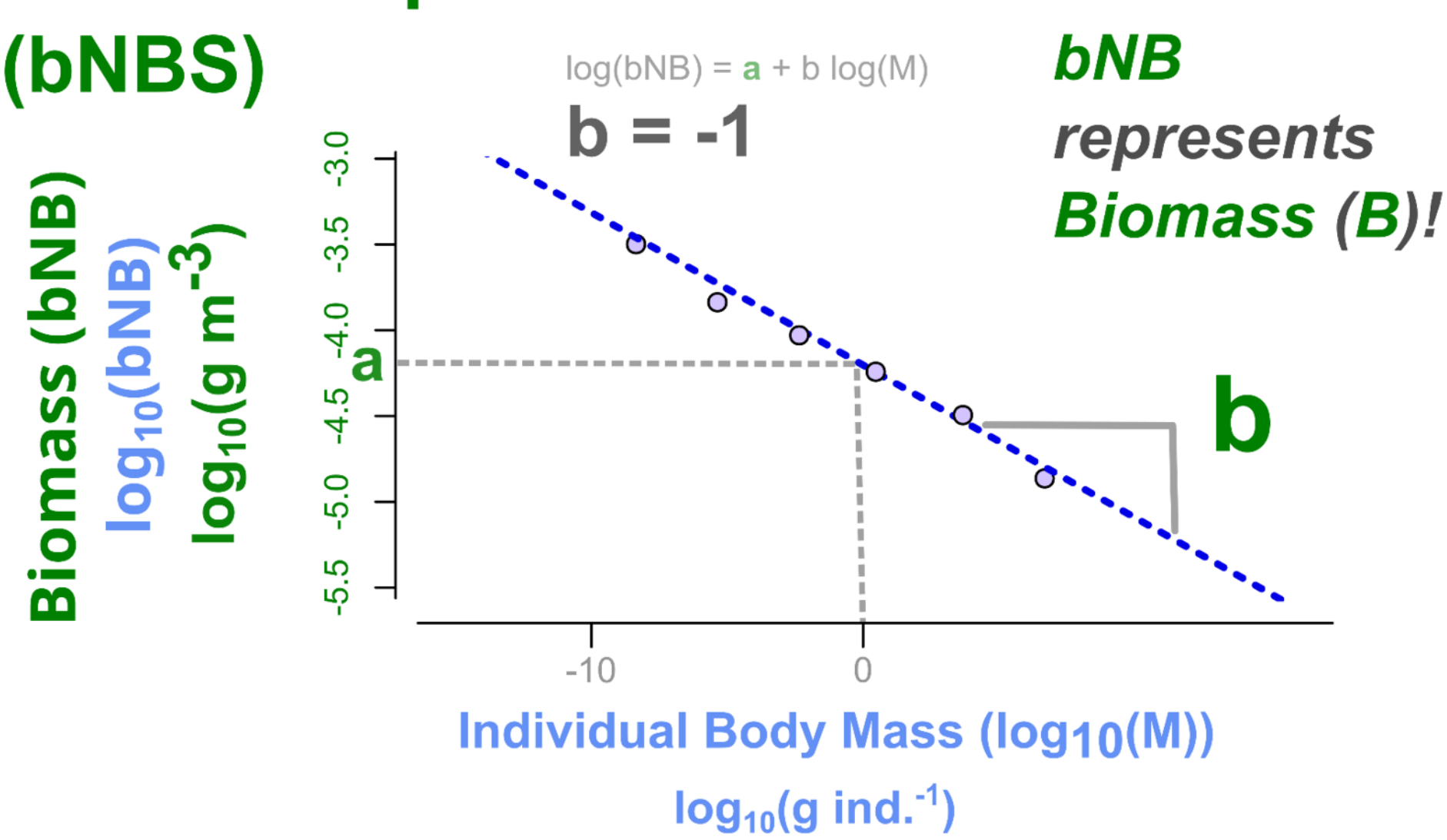

**Figure 4: Newly proposed bNBS method.** bNB: "backtransformed and normalized biomass" (the proposed variable). bNBS: "backtransformed and normalized biomass spectrum" (the proposed method, plot, and model). Note the changes in y-axis values from NBSS to bNBS. bNBS has the original biomass units.

## 2. Methods

### *2.1 Definitions of key parameters and their units*

Since the early days of quantitative ecology, there has been a wide range of confusing and ill-defined terminologies, specifically involving the term "density", which is still often used as a synonym for Abundance (ind. $m^{-3}$). In recent years, however, many authors have rather avoided using the term "density", to avoid possible confusion with other uses of this ubiquitous term. For example, there are many biogeochemical and plankton studies that use "density" as a synonym for mass-specific density i.e., the weight-to-volume ratio of particles (which is related to sinking rates). Another common use of "density" in ecology is the "kernel density" (Rosenblatt, 1956), which refers to a statistical method for estimating the probability density of observations (Sheather and Jones, 1991).

Therefore, to avoid confusion with many "density" meanings used by previous authors, and with the body-mass-specific density function ($D_x(M)$) defined further below, we are using the simple terms "Abundance" and "Biomass" through this study: **Abundance "A"** (e.g., ind. $m^{-3}$) simply refers to the total number of living individuals per space unit (area or volume), and **Biomass B** (e.g., g $m^{-3}$) simply refers to the total living mass per space unit (area or volume).

One might argue that, because A and B are functions of the sampled upper and lower body-size limits (i.e., of the mean body mass and the body-mass range), their units should not simply be expressed as numbers (or mass) per unit space, but rather in complex and quite awkward units, that would also account for the body mass range and for mean body mass (abundance: ind. $m^{-3}$ $g^{-1}$ $g^{-1}$, or: ind. $m^{-3}$ $g^{-2}$; biomass: g $m^{-3}$ $g^{-1}$ $g^{-1}$ or: g $m^{-3}$ $g^{-2}$, or: $m^{-3}$ $g^{-1}$). However, from the standpoint of dimensional consistency, we should never adopt such awkward complex units for A and B, since the units of any quantity should correspond to the dimensions of the physical quantity being measured, regardless of the factors that influence its numerical value. For example, the speed of a motorized vehicle is a function of its engine power (e.g., horsepower), but speed is still expressed in units of distance per time, not distance per time per horsepower. The latter unit would represent a different, more complex parameter (i.e., the change in speed per unit of engine power) rather than speed itself. Similarly, although abundance (and biomass) actually depends on the body-size range and mean body mass considered, the units should always remain simply numbers (or mass) per unit space.

Similarly simple units should also be used for cumulative quantities that account for lower and upper body-mass limits, such as the cumulative functions ($C_x$) of A and B. The **Cumulative Abundance Function $C_A$**(M) describes the increase in total abundance within a given body mass interval as the upper body mass limit M increases (Fig. 5). Accordingly, the **Cumulative Biomass Function $C_B$**(M) (Fig. 5) describes the corresponding increase in biomass within a given body mass interval as the upper body mass limit M increases. Both cumulative functions are calculated using a fixed lower body mass limit, $M_{lower}$, and an increasing upper body mass limit, $M_{upper}$. As $M_{upper}$ increases, both cumulative functions necessarily increase monotonically (Fig. 5). Although these cumulative quantities may appear more complex because they account for a range of body masses, each individual value of $C_A$ and $C_B$ nevertheless retains the same units as A and B, respectively: numbers per unit space for $C_A$ and mass per unit space for $C_B$. The body-mass limits define the range over which these quantities are accumulated, but they do not introduce additional dimensions into the units of the resulting cumulative quantities.

Next, we use the cumulative functions ($C_x$(M)) to define two body-mass-specific density ($D_x$(M)) functions: the **Abundance Density Function $D_A$(M)** (ind. $m^{-3}$ $g^{-1}$) is defined as the function that represents the rate of change (i.e., the first derivative) of the cumulative abundance distribution $C_A$ ($m^{-3}$) with body mass M (g $ind^{-1}$) (Fig. 5). Analogously, the **Biomass Density Function $D_B(M)$** (g $m^{-3}$ $g^{-1}$, or simply: $m^{-3}$) is the rate of change (i.e., the first derivative) of the cumulative biomass distribution $C_B$ with body mass M (g $ind^{-1}$) (Fig. 5).

Cumulative functions and density functions both actually constitute a description of the relationship of $C_x$ and $D_x$ (both being meaningful original dimensions in ecology), with body mass. This rationale and terminology follows the ideas presented by Blanco et al. (1994), with minor modifications.

Below is a short list of key definitions and abbreviations (in alphabetical order):

A: Abundance (ind. $m^{-3}$)

B: Biomass (g $m^{-3}$)

BWIB: "bin-width-inflated biomass" (g $m^{-3}$), i.e., the biomass-per-bin dataset generated by non-linear binning, prior to normalization.

bNB: backtransformed normalized biomass (g $m^{-3}$)

bNN: backtransformed normalized numbers (ind. $m^{-3}$)

bNBS: backtransformed normalized biomass spectrum (i.e., the plot and linear model that describes the relationship of $bNB_i$ *vs* $M_i$)

bNNS: backtransformed normalized numbers spectrum (i.e., the plot and linear model that describes the relationship of $bNN_i$ *vs* $M_i$)

$C_A(M)$: Cumulative Abundance function (ind. $m^{-3}$)

$C_B(M)$: Cumulative Biomass function (g $m^{-3}$)

$D_A(M)$: Abundance Density function (ind. $m^{-3}$ $g^{-1}$)

$D_B(M)$: Biomass Density function (g $m^{-3}$ $g^{-1}$)

M: Body mass (g)

$M_i$: Body mass (g) midpoint of the interval "i" (i.e., the geometric mean of body size in interval "i" in linear space, which corresponds to the euclidean midpoint in log space).

NBSS: Normalized biomass size spectrum (g $m^{-3}$ $g^{-1}$)

NNSS: Normalized numbers size spectrum (ind. $m^{-3}$ $g^{-1}$)

$w_i$: body mass (g) range (i.e., bin width) of the interval "i"

*2.2 Checking for issues in the NBSS plot method*

Extensive simulations with synthetic and *in situ* data were used to verify, in a careful step-by-step analysis, the procedures involved in transformations that lead to the popular NBSS plots, and to compare the original biomass data with the binned outputs.

Extreme bins (lower and upper edges) are often affected by border effects (lower y values than in the original distribution) and artifacts of empty bins (increased y values in the adjacent non-empty bins). Accordingly, all analyses and equations considered only the most "data-rich" central size range, where the original ecosystem size spectrum is appropriately represented by the binning vector, without considering the upper and lower edges of the spectrum (Suppl. Mat. Fig. SM1b), or size ranges with empty bins (Schwamborn et al., 2025).

For the *in situ* data zooplankton data, a selection strategy was applied to detect and select the linearly down-trending part of the size spectrum in any given dataset (Suppl. Mat. Fig. SM5), from the maximum to the first empty bin. Additionally, a sampling selectivity factor may be necessary for highly vagile organisms (e.g., net-caught zooplankton, or fish caught in towed nets), which is a unitless fraction. For the sake of simplicity, we are not addressing the topic of linear and non-linear sampling factors here, since such dimensionless factors do not change any units. and dimensions, which are the focus of this study. Linear models were fitted to log-transformed data by robust regression (less prone to outlier effects, than common OLSR, see Figs. Suppl. Mat. SM 6 a,b), using the rlm() function in the MASS R package (Venables and Ripley, 2002).

For synthetic data, histograms (linear binning and NBSS) were reconstructed from the original power-law shaped density distributions with known power law parameter “lambda” ($B = C * M^{-lambda}$). To evaluate the accuracy of normalized size spectrum methods (e.g., NBSS and bNBS) linear model log-log linear slope “b” estimates ($b_{estimate}$) were compared to the input biomass-body mass power-law exponent “b” ($b_{input}$) of synthetic data, where $b_{input} = -1 * (lambda -1)$; lambda is the power-law numbers-body mass distribution exponent $N \sim M^{lambda}$. Percent bias (PB) was then calculated as $PB = (b_{input} - b_{estimate}) / b_{input} * 100$. Bias levels were *a priori* defined (Schwamborn, 2018) as “negligible bias” (<5%), “reasonably accurate” (5 to 10%), “moderate bias” (10 to 30%), and (4) “severe bias” (> 30%). Also, I examined the relationship (correlation) between the estimated intercepts (a) and slopes (b) under numerous settings and variables.

To verify whether the variability in the original data is consistently represented in the output data and models, a non-dimensional coefficient of variation of residuals (CVR, %) was calculated for BWIB (bin-width-inflated biomass), NNSS (normalized numbers size spectrum), and NBSS, for all simulations. CVR (%) was calculated as the standard error of the residuals (SER) of the log-log-linear model (fitted with ordinary linear regression), divided by the mean of predicted "y" values, and multiplied by 100: CVR(%) = (SER / mean) * 100. The CVR-to-slope-ratio (i.e., error / information/ or noise / signal ratio, or variability-to-slope-ratio, VSR) was then computed and compared across multiple methods (e.g., NNSS *vs*NBSS, direct *vs* indirect biomass binning).

Bias (%) in normalized size spectrum methods (e.g., NBSS and bNBS), regarding slope "b" estimates and variability estimates (VSR) were compared across input values, linear and nonlinear binning methods, abundance and biomass and direct vs indirect (B = A * M) biomass binning methods. For each comparison, a total of n = 2000 simulations with different input values (with $b_{input}$ ranging from -1.6 to -0.4, representing a range from extremely steep to flat biomass spectra) were used. Numerous different input variability settings (from perfectly log-log-linear spectra towards "noisy" and "bumpy" high-variability spectra, dotted with many discrete peaks and domes) were tested. A total of n = 100.000 body mass values were generated and analyzed in each simulation. Thus, several hundreds of millions of body mass values were simulated in this study.

Possible binning effects were investigated with linear binning (constant bin width w, Suppl Mat., Fig. SM1a), and with non-linear, logarithmic binning (geometrically increasing bin midpoints $p_i$ and bin widths $w_i$). Commonly used logarithmic binning vectors in NBSS research have equal width in log space, where

$$w_i = w_0 * z_i$$

Common z values used in the literature for logarithmic binning vectors include z = 0.301, 1.5, and 2. For instance, z = 2 means that each successive bin is twice as wide as the previous one within a log2-scaled binning vector (e.g., Fock et al., 2026; Schwamborn et al., submitted). All binning vectors are "right = TRUE" (the default in R), where the histogram intervals are right-closed and left-open, meaning that each bin includes its upper boundary but excludes its lower boundary.

Additionally to numerous synthetic data, I also used datasets obtained *in situ*, i.e., zooplankton body mass and biomass (g C ind.$^{-1}$ and g C m$^{-3}$) data from the western tropical Atlantic, that were obtained from the TRIATLAS size spectra database (Fock et al. 2024, Fock et al. 2026, Schwamborn *et al.*, submitted). Since the underlying distribution parameters of these communities are *a priori* unknown, for such *in situ* datasets, only the linearly binned, Kernel density estimation (KDE, see below), and NBSS (logarithmic binning) results were compared, to verify the correctness and possible distortions due to the NBSS calculation methods.

Also, a thorough literature analysis was conducted regarding common units and plots, and intercept and slope values, in NBSS-based ecosystem science, using common databases and software (e.g., Google Scholar, Scopus, ISI Web of Science, PSSdb size spectra database: https://zenodo.org/records/11050013, GLOSSAQUA size spectra database: https://zenodo.org/records/14701391, and the TRIATLAS size spectra database: https://zenodo.org/records/13627093), with the main objective to verify the use of abundance units (e.g., "ind. m$^{-3}$", "(g / m$^3$ ) / (g / indiv.)", or "ind. m$^{-2}$") in NBSS plots.

*2.3 Kernel density estimation of size spectra*

Kernel density estimation (KDE, a non-parametric way to estimate the probability density function, Rosenblatt, 1956, Sheather and Jones, 1991) was applied to the raw body mass data, as a simple and intuitive means to visualize the shape of the size spectrum (e.g., peaks, bumps, and throughs), additionally to binning methods (e.. NNSS). KDE was applied using the "density" function in R, testing different smoothness levels, by varying the bandwidth value (e.g. bandwidth = 0.2 and 0.25), by using the R default Silverman's 'rule of thumb' (Silverman, 1986), and by using the Sheather-Jones bandwidth choice algorithm (bandwidth = "SJ") for heavy-tailed distributions (Sheather and Jones, 1991). To compare the results of KDE to different binning schemes, a "kernel-vs-binning median difference" (KBMD) bias estimate was calculated across all bins, after calculating an abundance estimate by binning and by KDE, for each bin. KDE can only be used for cross-checking abundance spectra (NNSS), not for NBSS, since kernel density only applies to abundance (counts are converted into kernel density). Results of transformations based on normalized non-linear binning (i.e., NNSS) were considered realistic representations of the original data, if they matched the KDE results (Fig. 6).

*2. 4 Developing and testing the new "backtransformed" bNBS method*

Based on the observation that the currently used NBSS method does not adequately represent biomass (having wrongful dimensions, units, and numerical values), a new method was developed in a careful step-by-step analysis and iteration procedure (see Results), and subsequently tested.

**Table 1** Detected issues and proposed solutions in normalized size spectra research

| **Detected Issues** | **Proposed Solutions** |
|---|---|
| *1. Inadequate ("normalized") units* | 1. Backtransformation *to original, quantitative units after normalization (bNNS a*nd bNBS) |
| *2. Border effects* | 2 Use only the most data-rich section of the spectrum. Exclude borders and sections with empty bins. Use standardized selection algorithms (e.g. maximum to first non-empty bin) |
| *3. Binning artifacts (all binned methods) hamper the detection of peaks and throughs in the spectrum, leading to oversimplification and linearization.* | 3. Kernel density estimation (KDE) with optimized bandwidths. |
| *4. linearizing artifacts in indirect NBSS (B = A * M)* | 4. Use direct binning for biomass (do not calculate B through binned abundance, such as B = A * M) |
| *5. Ambiguous, imprecise, and incorrect "size spectrum" terminology.* For instance, "NBSS" contains the wording "size", which implies *body size* (not body mass). Most NBSS studies deem synonymous 1.) the name of the size spectrum and model and 2.) the name of the normalized biomass variable. | 5. Proposed terminology:<br>The biomass-body-mass spectrum (plot and linear model): bNBS plot, bNBS model, bNBS slope, bNBS shape.<br>Variable: bNB<br>Title of the y axis: "bNB, $log_{10}(g\ m^{-3})$",<br>or simply "Biomass, $log_{10}(g\ m^{-3})$"<br>Text: "bNB values", or simply "biomass values".<br>Also applies to bNNS ("backtransformed normalized numbers spectrum", i.e., the abundance-body-mass spectrum). |
| 6. Strong correlation of intercept "a" and slope "b", impeding naïve analyses of both descriptors. | 6. Use a slope-independent estimator of biomass (e.g., total predicted B, or mean B, or B at selected bins) |
| 7. Impossibility of quantitative comparisons of abundance or biomass across studies, regions and time periods, with current NBSS and NNSS. | 7. For bNBS, use the "breaks" (bin edges, right-closed) of the exact standard $log_2$ binning vector. Always inform body mass range and exact bin width when informing abundance or biomass values at a specific body mass (e.g. at M= 1g ⇔ log(M) = 0). Use a standardized factor to convert from NBSS to bNBS (e.g., a unity reference bin with wR = 1g) |

Considering that linearly binned biomass histograms represent and preserve the original biomass-scaled numerical values, dimensions, units, and total biomass ($B_{total}$), but the currently used NBSS method does not, the desired new method should be a reliable and simple way to obtain correct numerical values (e.g., total Biomass) and represent a relevant improvement regarding dimensions and units, while retaining the advantages of the popular NBSS method (intuitive linear shape, straightforward linear model fit, and convenient non-linear binning). Thus, log-log linear slope "b" estimates, total biomass $B_{total}$ estimates, units, and plots of the new, herein proposed "backtransformed" normalized biomass spectrum (bNBS, Fig. 4) method were compared with linearly binned and standard NBSS outputs, for synthetic and *in situ* data.

*2.5 Accounting for the structure of the original data - the multi-step bootstrap*

Common binning-based linear regression (as used in NBSS studies) uses the degrees of freedom of binned data, ignoring the original sample sizes. Thus, a new multi-step bootstrapping routine (as in Schwamborn et al., 2019) was developed and used for the calculation of proper confidence intervals, with 10,000 resampling runs for each step. First, resampling (sampling with replacement) was conducted on the original body mass data (step1 ), prior to binning. Then, the bins (pairs of counts and mids) were resampled within the linear regression procedure (step 2). Simply put, the routine consists in bootstrapping the raw body mass data, binning the data, normalizing, fitting the linear regression in log–log space, storing intercept and slope each time, and finally, computing the 95% confidence envelopes for plotting.

The 95% confidence intervals for each parameter estimate were obtained as the 95% quantiles of the bootstrap posterior distribution, for each of the relevant size spectra parameters (e.g., intercept, mean predicted “y” value, and slope of the log-log-linear model).

Detailed descriptions of the new robust, bootstrapped, back-transformed bNBS method, including all data and simulations, are available at github.com/rschwamborn/bNBS.

## 3. Results

### *3.1 Biomass spectra are scaled as biomass (not abundance or biomass flux)*

In all simulations, the scaling (i.e, the log-linear slope) of the common NBSS plot (e.g., its linear model slope and potential bumps and troughs) was a faithful, highly precise and accurate representation of the scaling of the original biomass-body-mass distribution (Fig. 6). Bias in NBSS slope "b" was always negligible, always with less than 2% bias (mostly less than 0.7 % bias), when using non-linear binning and at least 5 bins (comparing the scaling of input data *vs* fitted log-log linear model slopes, synthetic data, n = 2000 simulations with input scaling $b_{biomass\text{-}body\text{-}mass}$ values from -1.6 to -0.4). Thus, all simulations proved that the NBSS plot is an adequate index (i.e., has the same slope and shape) of the original biomass distribution, not abundance or biomass flux (“dB/dM”). If NBSS represented biomass flux (dB/dM), its scaling would necessarily differ from that of the original biomass-body-mass distribution, because (dB/dM) is the first derivative of biomass. Thus, we must reject the "NBSS-is-biomass-flux" hypothesis, and accept the "NBSS-is-biomass" hypothesis.

Yet, all reviewed literature used some form of biomass flux units (e.g., "(g / $m^3$ ) / (g / indiv.)" ) or abundance units (e.g., ind. $m^{-3}$, or ind. $m^{-2}$) in their NBSS plots, reflecting the calculation procedure to obtain the NBSS data (i.e. the division of biomass by bin width). Also, all reviewed NBSS studies confounded the terms for the size spectrum plot (the name for the plot, the model and the y vs x relationship) with the normalized biomass variable (the "y" values in the plot), using the same term and wording ("NBSS") for the plot, the log-linear model, and the normalized variable. None of the reviewed literature informed whether direct (sum of body mass per bin) or indirect binning (through total abundance per bin and body mass at geometric mean bin midpoints, where B = A * M) of biomass was conducted, to obtain the NBSS.

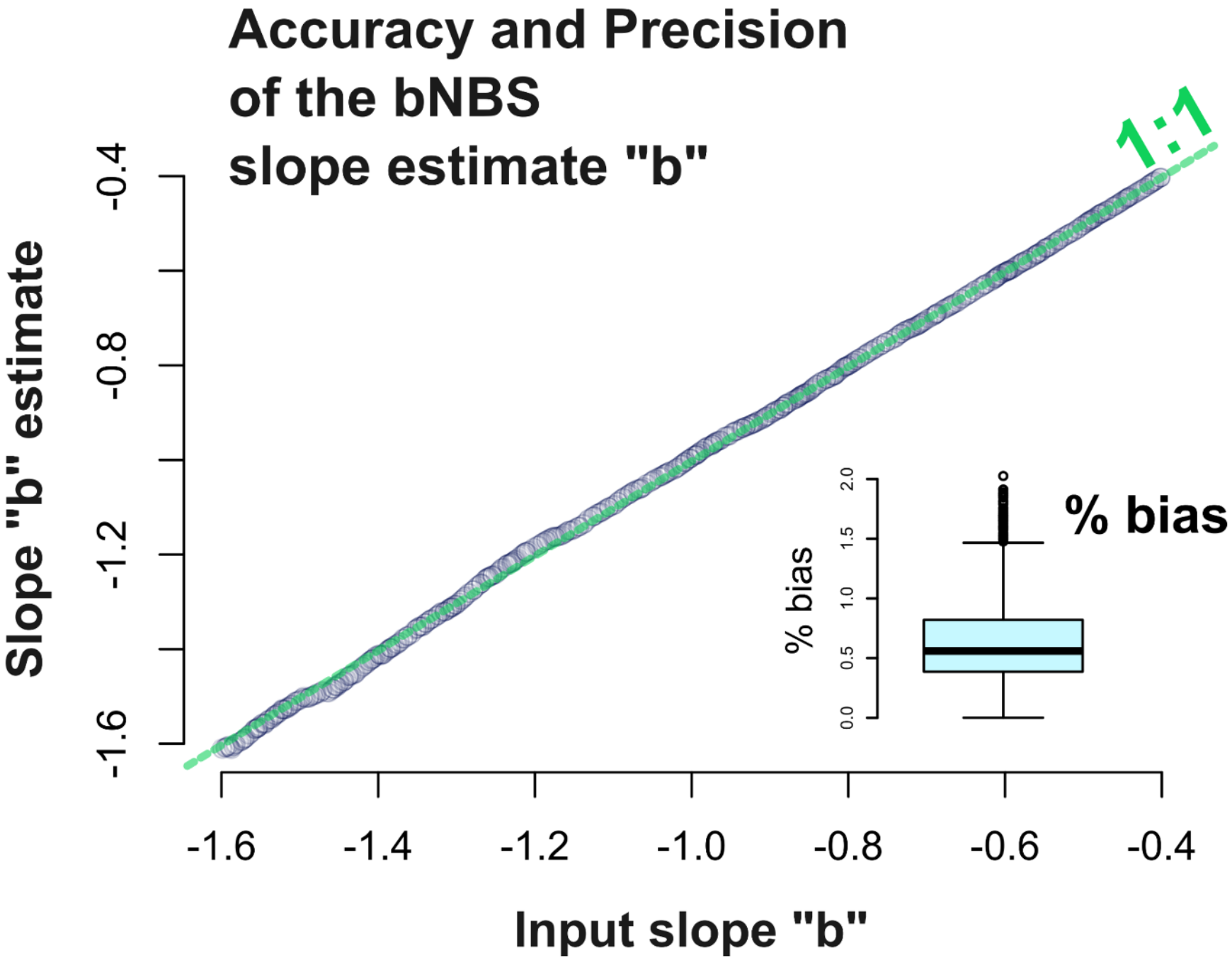

**Fig 5.** Accuracy and precision in the estimation of log-log-linear model slope "b" for bNBS ($b_{estimate}$ *vs* $b_{input}$). N = 2,000 simulations with n = 100,000 individuals per simulation. Based on 2,000 near-perfect (low - variability) size spectra with known, exact input slope values $b_{input}$. Calculation with direct biomass estimations (i.e., based on the direct sum of all body mass values for each bin), using method S1 ("k-normalization"). Inlet: percent bias. Note that bNBS is an excellent estimator of the ecosystem size spectrum slope, i.e., it shows only negligible bias, with less than 2% bias (mostly less than 0.7% bias).

Another issue observed in most reviewed NBSS literature was that intercepts (a) and slopes (b) were treated as if they were two useful, independent estimators (i.e., two relevant ecosystem properties). Yet, in any analysis involving a matrix of linear models, a and b estimates are always statistically correlated, except in the special case where the x vector is centered such that $x = 0$ is located at its midpoint (i.e., in the unique case where when "M = 1 g" is at the exact center of the NBSS data used for linear model fitting).

After examining the relationship between the estimated intercepts ("a") and slopes ("b") under numerous settings and variables, in all cases, there was an almost perfect linear relationship between a and b, with $R^2$ values of approx. 1, and highly significant p-values (Suppl Mat., Fig SM x relationship between a and b ). This occurs because, in the x vector used for our simulations, the reference value ($x = 0$) is located far from the midpoint of the x vector (specifically, near its upper extreme). Under such conditions, intercept estimates are always algebraically and statistically tied to slope estimates, impeding any naive analysis of both descriptors. Therefore, any analyses investigating the effects of external variables (e.g., temperature) on both parameters (a and b), under such settings, have to explicitly account for their complex covariance matrix structure. And obviously, it is not possible to construct a linear model based on the plot of a *vs* b (Suppl Mat., Fig. SM2 relationship between a and b), a "textbook" case of spurious autocorrelation i.e., a model based on two non-independent variables.

However, there is no inherent reason to use the intercept (i.e., the predicted value at $x = 0$) as the descriptor of the vertical position of any linear model. The intercept depends on an arbitrary choice of $x = 0$ (e.g., biomass at x = 1 g, $\log(x) = 0$, which depends on the arbitrary choice of units, e.g., 1 g, or 1 mg), and therefore may not provide any meaningful measure of model elevation (i.e., ecosystem biomass). A more robust approach for the interpretation of variations in both a and b (and their responses to extrinsic ambient variables) is to replace the intercept with a quantity that does not depend on the arbitrary positioning of x=0, and on the slope “b”. Two simple, robust alternatives to obtain ecosystem biomass estimates from linear models are: i.) use the predicted y value at the midpoint (or mean) of x, or: ii.) use the sum (or mean, or median) of predicted y values (or observed y values) across all x values used for model fitting. Alternatively, one may consider using the observed (not the predicted) y values (for example the sum of the observed y values, i.e., total abundance, or total biomass), instead of a linear model property, to assess ecosystem biomass.

Due to the normalization, in NBSS (and NNSS), the slope and vertical position (mean y value) are independent of the binning vector used (a big advantage of this method). Yet, the exact value of the intercept (i.e., y at log(x) = 0), depends on the units of y and on the units of x (gC vs mg C or pg C, ml vs $mm^3$, etc. will define the exact position of log(x) = 0, and thus the numerical value of the intercept). A standardized binning vector, with standard units, is urgently necessary for NBSS, as to be able to compare the biomass (or abundance) among natural ecosystems.

### *3.2 Loss of variability by binning*

First, it is important to remember that all histograms for size spectra analysis are based on the numbers of individuals (where each observation or count is one individual), not samples. In any frequency distribution histogram, frequency may be given as a fraction of 1 ($F_1$, or total density, where the total area of the histogram is 1, as in KDE) or as absolute counts of observations or samples ($F_{abs}$, as used throughout this study, except for KDE).

In a frequency distribution histogram of individuals (not samples) with absolute counts (not relative frequency), the y-axis value of each bar represents the total number "N" of individuals in each size class, which can be directly converted to Abundance A, with A = N/S (dividing by a standard space unit S, i.e. of area or volume, e.g., after dividing by 1 cubic metre). When subsequently multiplying Abundance A by individual mass M (i.e., B = "height * midpoint" in each histogram bar), Biomass estimates can be obtained indirectly for each size class.

Direct pooling of biomass per bin (i.e., obtaining B directly for each bin, instead of first pooling for A for each bin i, and then converting $A_i$ to $B_i$ by $B_i = A_i * M_i$), produced exactly the same estimate for the log-log-slope "b", in all simulations. However, our results confirmed that it is conceptually better (less autocorrelation and a better representation of the variance in M), to sum all individual body mass values per size range (calculating Bi directly, by summing all M values in each bin), instead of first obtaining $A_i$ and then converting into $B_i$ (directly estimating $B_i$ also avoids potential errors associated with calculating and applying the geometric mean body-mass midpoint for each bin). This rationale was confirmed by the significantly better representation of variability in body mass by the direct binning method, than when using the indirect ("$B_i = A_i * M_i$") method (Suppl. Mat., Figures SM3 and SM4).

The variability represented in the spectra (calculated as the coefficient of variation of residuals, CVR, %) was always considerably and significantly ($p < 10^{-16}$) lower (generally more than twice as high) in abundance spectra (NNSS, normalized numbers size spectrum) than in biomass spectra, indicating that the transformation from abundance into biomass does not have a variance - reducing (i.e., linearizing SAA) effect. Rather (as expected for a meaningful calculation), body mass introduces additional information and variance. Even when accounting for the difference in slope between abundance and biomass (i.e., $b_{NBSS} = b_{NNSS} +1$ ; abundance spectra are generally approximately twice as steep, and thus, the "y" values are spread approximately twice as wide), the difference in variability represented in NNSS and NBSS was still highly significant (higher variability of NBSS than of NNSS) , as observed in all variability scenarios (variability-to-slope-ratio, VSR, $p < 10^{-16}$).

Regarding the comparison of biomass spectra ($NBSS_{indirect}$) obtained indirectly, from binned abundance (B = A * M) or by directly binning the original body mass data ($NBSS_{direct}$), there were no significant or relevant differences in slope (near-zero bias in slope "b" estimates). However, there was a highly significant ($p < 0.0001$, permutation test for medians, n = 2000 simulations) difference in variability estimates (i.e., in variability-to-slope-ratio, VSR), with the direct binning method presenting higher VSR estimates than the indirect biomass estimation method. This was observed in all variability scenarios (e.g., low variability scenario: VSR with indirect Biomass calculation: median = 0.152, VSR with direct binning : median = 0.166; $p < 0.0001$, bias in VSR: 9.2 % ; high variability scenario: VSR with indirect Biomass calculation: median = 2.37, VSR with direct binning: median = 2.44; $p < 0.0001$, bias in VSR: 3.2 %).

Thus, these simulations showed the absence of any relevant statistical artifact (SAA) effects on slope "b" estimates (and in variability estimates, when using direct binning), confirming the validity, correctness, precision and accuracy of the NBSS approach. The slightly (3.2 to 9.2 percent bias) higher variability displayed in the direct method Suppl. Mat., Figures SM3 and SM4) indicates that it is superior to the indirect method (i.e., less prone to SAA linearizing effects), for representing the variability in the original body mass data.

*3.3 Abundance and biomass are functions of the sampled size range*

Starting from simple, first principles, let's look at the relationship between individual body mass and abundance. It is obvious that the number N of individuals caught in a sample is a function of mean organism abundance A (ind. $m^{-3}$) in nature, and of the chosen reference space unit (S), i.e., sampling volume (e.g., 1 $m^{-3}$).

Let's first consider a hypothetical (very unrealistic) scenario where all organisms have exactly the same size: in this scenario of zero size variance, body size can be ignored in estimating abundance. The number of organisms N in any given sample, in such a simple scenario, can be described as a function of Abundance A (ind. m-3) A = N / S, where N has units of individuals.

When considering that size (and body mass) in nature is variable among individuals, we must acknowledge that the number of individuals captured in any sample depends on the size range (or body mass range "w", in g) considered. A wider body mass range "w" (maximum - minimum individual mass in the sampled body mass interval, in g) will obviously yield a larger number of sampled organisms. Furthermore, in a power-law-distributed pelagic community, abundance is a function of body mass M (i.e., the geometric mean body mass in a given size range, in g), with larger organisms being less abundant. Accordingly, abundance, as a function of M and w, can be expressed as A(M,w) = $D_A$(M) * w(M), where $D_A$(M) is the mean size-specific abundance density distribution function (ind. $m^{-3}$ $g^{-1}$).

The $D_A$ (M) function (ind. $m^{-3}$ $g^{-1}$) is derived from the cumulative abundance $C_A$ (ind. $m^{-3}$), i.e, it is the first derivative of the cumulative abundance distribution function $C_A$(M). Thus, the $D_A$(M) function describes how fast cumulative abundance ($C_A$) accumulates with increasing body mass. In a common pelagic food web, the exponent of the $D_A$ *vs* M relationship is -2, while the $C_A$ function (the integral of the $D_A$ function) has an exponent of -1 (Fig. 5). The $D_A$(M) function is well represented by the common NNSS (normalized numbers size spectrum, Blanco et al., 1994, Vandromme et al., 2012) slope of approx. $b_{NNSS}$ = -2 (Figueiredo et al., 2025).

Accordingly, the biomass density distribution function $D_B$(M) represents the relationship of of biomass density $D_B$ (g $m^{-3}$ $g^{-1}$, or simply: $m^{-3}$) and body mass M. Herby, $D_B$ is calculated as $D_B$ = d($C_B$) / d(M). Thus, the biomass density distribution function $D_B$ (M) is the first derivative of the cumulative biomass distribution function $C_B$ (M). Analogously, the $D_B$ (M) function describes how fast cumulative biomass $C_B$ (g $m^{-3}$ ) accumulates with body mass. The $D_B$(M) function is well represented by the NBSS, with a slope of $b_{NBSS}$ = -1 (Dugenne et al., 2024). Interestingly, a log-linear slope of exactly b = -1 in the $D_B$(M) function leads to a logarithmic integral in the $C_B$ function, which is not a power-law distribution (Fig. 5).

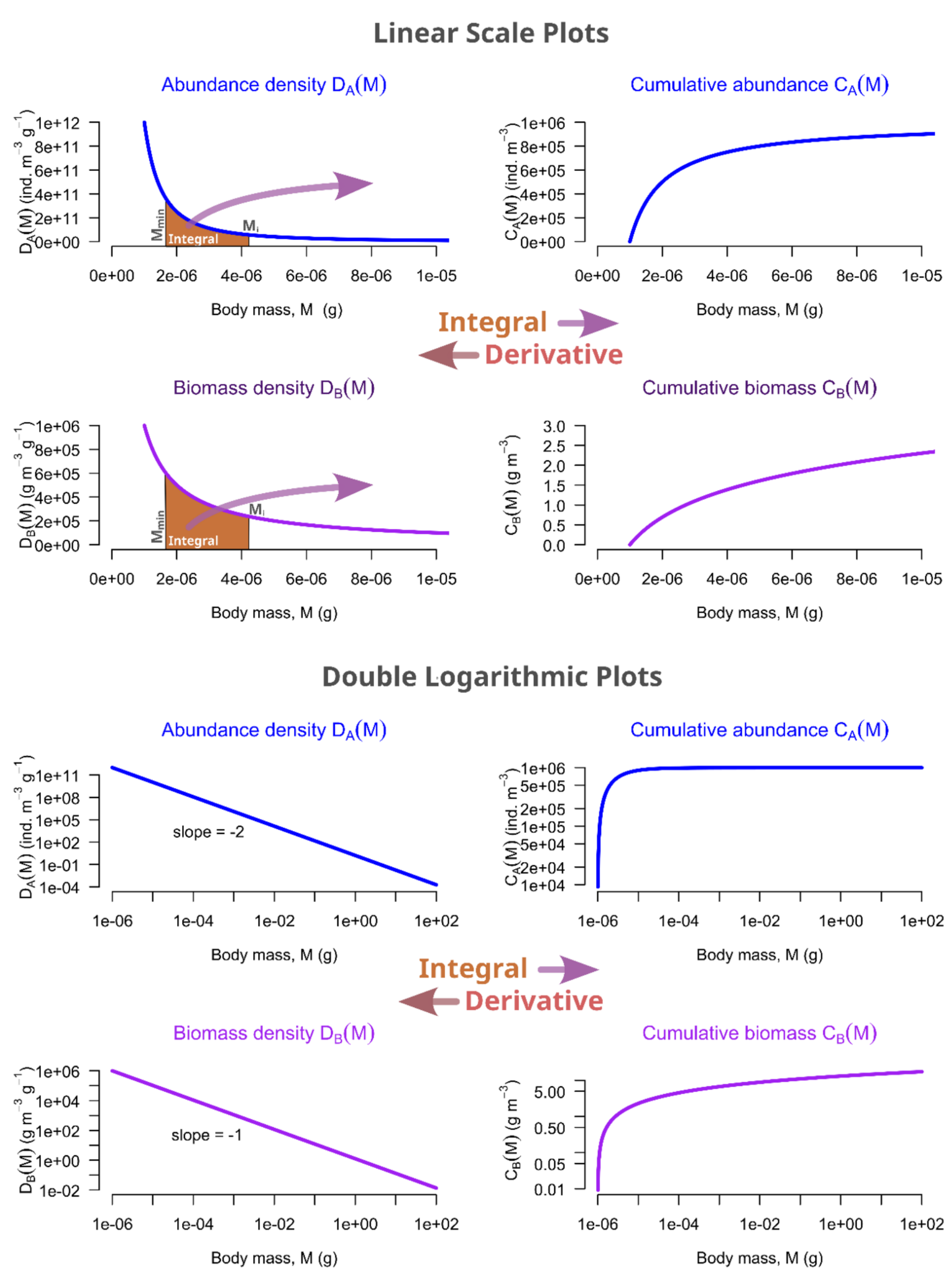

**Fig. 5.** Density distributions and their cumulative integrals. The abundance density distribution, $D_A$ (M) (ind. $m^{-3}$ $g^{-1}$), and the biomass density distribution, $D_B(M)$ (g $m^{-3}$ $g^{-1}$, equivalent to $m^{-3}$), are the first derivative of the cumulative abudance distribution function, $C_B(M)$ (ind. $m^{-3}$) and of the cumulative biomass distribution function, $C_B(M)$ (g $m^{-3}$), respectively. The cumulative functions ($C_x$) are the integral of the density (Dx) functions. Each cumulative value ($C_x(M_i)$) is obtained by integrating $D_x(M)$ from a fixed lower body-mass limit ($M_{min}$) to a variable upper limit ($M_i$). Upper panels: linear scale plots; Lower panels: double logarithmic plots

### *3.4 Binning transformations - the relationship between binned and original biomass*

Considering that NBSS does not represent abundance or biomass flux (see 3.1), but still has awkward, non-intuitive abundance units (such as "pseudo-abundance" units, or "biomass flux" units), it becomes evidently necessary to develop a new calculation procedure, that produces adequate biomass units, thus solving the "NBSS-Biomass-Abundance-paradox"

Thus, the main objective of this effort became to find a back-transformation method that transforms NBSS values back into the original biomass units, dimensions, and values. For this purpose, we will first looked at how the binning process distorts and transforms the original biomass into the binned biomass, or "bin-with-inflated biomass" (BWIB), which is the main transformation occurring at the onset of the NBSS plotting method. Below, we will consider several possible scenarios of binning vectors and density distributions of the original population, to investigate the relationship between original, non-transformed (i.e., linearly binned) biomass $B_i$ and transformed (i.e., non-linearly binned) biomass $BWIB_i$.

### *3.5 Linear vs non-linear binning*

For the most simple case, with a constant (linear) binning vector, and a linear uniform (flat) distribution, the bin "y" value of each histogram bar (total Biomass per bin, Bi, estimated in each bin) and integral area values in each histogram bar scale linearly with bin width “w” used throughout the histogram, and with the inverse of binning interval range $R^{-1}$ (i.e., the interval from the smallest lower to largest upper value in the binning vector $R = M_{max,upper} - M_{min,lower}$).

Within size spectrum science, it is common to normalize abundance (“normalized numbers size spectrum” NNSS, Blanco et al., 1994, Vandromme et al., 2012) and biomass (normalized biomass size spectrum, NBSS, Platt & Denman 1977), i.e. to divide each $y_i$ value by the its corresponding bin width $w_i$. Thus, one obtains the NBSS plot (Fig. 2), where the normalized biomass is: $NBSS_i = BWIB_i / w_i$

In the case where we use a linear binning vector (constant bin width), there is a single value of w, and all $NBSS_i$ values can be simply back-transformed from NBSS units into the original biomass units and dimensions by simply multiplying each $NBSS_i$ value by the global w value.

When using a non-linear binning vector (e.g., with a $log_2$-scaled bin width), the relationship between NBSS and B is non-linear, becomes bin-specific and must be calculated individually for each bin. Yet, total biomass $B_{total}$ is still the sum of all $B_i$ values across the sampled size range R, as in regularly spaced bins.

$$B_{total} = NBSS_i * w_i$$

If we want to compare the “bin-with-inflated biomass” (BWIB) of nonlinear bins and linear bins, we may look at three forms of calculating total biomass (across the complete size range): 1.) the initial total biomass (Btotal, prior to binning), 2.) the sum of all linearly binned $B_i$ values and 3.) the sum of all non-linearly binned $BWIB_i$ values. In all datasets, these total biomasses were obviously always numerically identical within each dataset. Thus, we may conclude that linear and nonlinear binning divides the same total B into different proportions, but the total biomass is always the same ($BWIB_{total} = B_{total}$). Thus, we can conclude that B and BWIB have exactly the same units and dimensions, of true biomass, and that the distortions in the creation of non-linear BWIB do insert information of the w vector into the data, but do not turn it into a different parameter (e.g., a "biomass * w" - parameter, with different units and dimensions, different from biomass). Consequently, we must accept the idea that BWIB is a biased representation of biomass (weighted by w), but not a dimensionally different form of biomass, and thus B and BWIB have exactly the same units.

At first sight it may seem that BWIB contains relevant information from the M vector, and thus, one may hypothesize that BWIB has different dimensions and units than the original biomass. Yet, the *dimensional invariance* of BWIB, as confirmed in this study (i.e. the equality of sums, $BWIB_{total} = B_{total}$), is a key finding to understand the nature of the transformations that occur during the NBSS method.

Thus, the current biomass units for BWIB (e.g., g $m^{-3}$) are correct. Accordingly, when dividing BWIB by w ($NBSS_i = BWIB_i / w_i$), there is an evident change in dimensions and units, and thus the current NBSS units (units of B / M) are also correct. Yet, our simulations showed (see 3.1, above) that NBSS is an index of biomass (!), not biomass flux. This blatant contradiction leads us to postulate the need of additional corrections, re-dimensionalization and rescaling of the NBSS.

Here, we need a short reflection on the fundamental correctness of the units of B (the base of all rationale regarding unit transformations): If B is dependent on background ecosystem biomass density D level and units (e.g., "$m^{-3}$"), and sampling units (e.g., cubic meter) only, it should be g $m^{-3}$. Yet, if we acknowledge that B is dependent also on mean body mass M and mass range (bin width w), maybe the correct units of biomass should be g $m^{-3}$ $g^{-1}$ $g^{-1}$? The hypothetical adjustment of fundamental dimensions and units of B would also affect the units of BWIB and NBSS and could, at first sight, be a simple solution to the apparently awkward units of NBSS. The short answer is that common units of B such as "g $m^{-3}$" are correct, since practical units simply reflect what is being measured physically (mass per volume). Changes in M and w indeed modulate the numerical value of B, but do not change its dimension and units. Consequently, the units of BWIB (g $m^{-3}$) and NBSS ($m^{-3}$) are also mathematically correct.

Simply put, the results and rationales above showed that the NBSS method is correct, but incomplete, needing an additional final corrective step (the back-transformation, i.e., back into units and dimensions of biomass).

*3.6 Back-transforming from NBSS to biomass in a "NBSS-is-biomass-flux" scenario*

Under a scenario where NBSS would represent biomass flux (not original biomass), back-transforming from NBSS to the original B units would have to be accomplished by integrating over body mass ($Bi = NBSS_i * w_i$). Integrating a function f(x) always leads to a radical change of units, values, dimensions, shape and slope of the plot (except for a perfect exponential function: $f(x) = e^x$), which is invariant under differentiation or integration). Thus, under the (unrealistic and already rejected, see 3.1 ) “NBSS-is-biomass-flux hypothesis”, the necessary back-transformation (i.e., correction) of NBSS units to the original B units would lead to a radical change in the shape of the biomass-body size spectrum (the integral of a b = -1 NBSS would actually have b = 0, which is not a power-law distribution), challenging all we know about natural ecosystems.

Conversely, in the following steps, considering the (already proven and accepted, see 3.1 above) “NBSS-is-biomass-hypothesis” we may simply (linearly) back-transform units and values from NBSS to B, without any need to integrate over body mass and to change the shape of the spectrum. Thus, we will indeed be able to correct the NBSS numerical values, while preserving current size spectra log-log-linear models (generally with b = -1), and all other insights and ecosystem models that are based on current ecosystem size spectra science.

*3.7 Back-transforming units and values in a "NBSS-is-biomass" scenario*

There are three possible solutions to the problems described above (Suppl. Mat., Figs SM 5a, SM 5b). Solution S1: retaining the original biomass units throughout the process of normalization (non-dimensional correction, by "k-vector-normalizing" the biomass), Solution S2: Back-transforming from the common NBSS to obtain the bNBS. Both solutions have an identical result: the "backtransformed" normalized biomass spectrum (bNBS). Solution S3: factor-normalization of the BWIB data (multiplying all values by a factor "f", that subtracts -1 from any non-positive BWIB slope), produces a hyper-distorted effect in peaks or domes with positive slope sections (not recommended, Suppl. Mat., Figs SM 5a, SM 5b).

*3.7.1 Solution S1: Non-dimensional bias correction with "k-normalization"*

In the (proven, see above) "NBSS-is-biomass" scenario, NBSS represents biomass (not the change of biomass with individual mass), and the misleading units of the currently used NBSS method must be corrected. We initially may try to find a way of retaining the original B units (e.g., g $m^{-3}$) throughout all calculations and transformations (Solution 1). Yet, when dividing $BWIB_i$ by bin width $w_i$, the resulting dimensional transformations cannot be ignored. Dividing biomass B by w (e.g., g indiv.-1) does indeed affect the original dimensions and units. The premises and calculations of the NBSS method are correct.

Yet, there may be a new way (instead of the current NBSS method) to normalize the bin-width-inflated biomass, while retaining the original B units and dimensions (Solution 1). It may be possible to use a new kind of normalization procedure: dividing BWIB by a unitless *relative bin width*, represented by the binning vector "k" (where $k_i = w_i / R_i$), to obtain the "k-normalized biomass" knB, within the following equation, for each bin "i": $knB_i = BWIB_i / k_i$

The log-log-plot and linear model of knB vs body mass M may be called the knBS ("k-normalized biomass spectrum") plot and model, a preliminary precursor form of the bNBS ("backtransformed normalized biomass spectrum").

The division by the unitless vector k is a non-dimensional bias correction and does not affect the original dimensions and units, then $B_i$, $BWIB_i$ and $knB_i$ all retain the original units of B (simply g $m^{-3}$, not units of abundance or biomass flux). This easily solves the paradoxes described above. Dividing by the relative bin with vector "k" (where $k_i = w_i / R$, and R is the overall range of the binning vector), instead of absolute bin with $w_i$ does avoid all the problems regarding the change in units and dimensions involved in the calculation of NBSS, since the vector k is dimensionless and unitless. However, due to the complex nonlinear interactions of the log2-binning vector and the power-law distributed data, normalizing with $k_i$ still does not achieve a perfect numerical scaling (identical values of $knB_{total}$ and $BWIB_{total}$). Thus, even if we use the vector k instead of w, we still have to conduct a final rescaling (correction) procedure, to obtain a correct representation of the biomass vs body mass spectrum in correct units and numerical values of actual biomass.

Thus, after redimensionalizing with k, rescaling is needed, by using the k-scaled correction factor F , where $F = BWIB_{total} / knB_{total}$

Each $knB_i$ value is then back-transformed into the scale of B, into a correctly scaled, fully "backtransformed" normalized biomass (bNB) by:

$$bNB_i = knB_i * F$$

Since the vector k and the factor F are both dimensionless and unitless, no units are changed throughout all calculations. Thus, the original biomass units (e.g. the units of BWIB and of B) are simply retained in the "backtransformed" normalized biomass spectrum (bNBS, e.g., g m-3, Fig. 4). In a final check, the sums of BWIB and bNB should always be identical ($BWIB_{total} = bNB_{total}$).

### *3.7.2 Solution S2: Back-transforming from NBSS to bNBS*

There are many situations where researchers have already conducted all transformations and calculations to obtain the hugely popular NBSS data, models, and plots, and may now wish to convert and correct them, i.e., to transform them from NBSS into bNBS. Using a slightly modified calculation (based on the rationale and calculations above), the desired backtransformation from NBSS to bNBS (i.e., back to the original biomass units, Fig. 4), can be easily achieved (Solution S2) within a simple rescaling and redimensionalization (i.e., linear correction) procedure.

Thus, after normalizing (dividing $BWIB_i$ by $w_i$, where $NBSSi = BWIB_i / w_i$) to obtain the common NBSS, rescaling is needed (Fig. 7). This can be done by using the w-scaled dimensional correction factor w' ("F-prime"), where

$$w' = BWIB_{total} / NBSS_{total}$$

All NBSS data are then transformed back (re-dimensionalized back and rescaled back) into the original scale, dimension, and units of biomass, i.e., into "backtransformed normalized biomass" (bNB), where

$$bNB_i = NBSS_i * w'$$

w′ can also be considered a type of biomass-weighted harmonic average bin width. Since w' thus has units of M (e.g., g $ind.^{-1}$), this calculation backtransforms the units of the NBSS into the original biomass units, within the bNBS (e.g., g $m^{-3}$). Similarly, the common abundance-based "normalized numbers size spectrum " (NNSS, Blanco et al., 1994, Vandromme et al., 2012, Figueiredo et al., 2025) can be easily converted to bNN (i.e., backtransformed into the original units of abundance, or "backtransformed normalized numbers") by multiplying NNSS with w'.

Multiplying by w' (Fig. 7) not only rescales the values (of NNSS and NBSS) numerically into the original scales, but also converts (i.e., redimensionalizes) their units back into the original units and dimensions (Suppl. Mat., Figs SM 5a, SM 5b). The fact that solutions S1 and S2 produce 100% identical results, represents a successful cross-check that proves the consistency of these two solutions. In solution S2, the single multiplicative constant w' restores biomass units and dimensions, while preserving the log-linear slope “b”, size spectrum shape (peaks and bumps), relative variability, and statistical regression results. One important modification is that multiplying by w' changes the log-scaled intercept “a” in a mathematically clean and intuitively simple way, from units of $D_B$ to units of B (Fig. 7).

Alternatively, many other potential values of w (e.g., the artimatc mean w, geometric mean w, or any other selected w value), instead of the biomass-weighted harmonic average bin width w’, may be tested and used for backtransforming from NBSS to bNBS, provided that the chosen global w value is clearly defined by the authors, within the methods descripton. A universally defined standard reference bin width value $w_R$ (e.g., $w_R$ = 1 g), may provide a standardized approach for backtransforming NBSS to bNBS across datadest (see Discussion below).

*3.7.3 Solution S3: factor-normalization of the BWIB data (subtracts -1 from the BWIB slope in non-positive sections)*

Solution 3 directly subtracts “1” from the BWIB slope, which avoids any possible linearizing normalization artifacts (dividing by a linearizing vector, such as “w” or “k”). The original variability in the BWIB data is retained, but the distorted (often near zero) slope of BWIB is corrected towards the real (often near -1) slope of the biomass-body-mass distribution.

Since linear model slopes behave in a multiplicative (not additive) way, we have first to find a factor that can be used for this transformation , i.e. compute a multiplier that subtracts -1 from the original BWIB slope. Thus, we first compute the non-dimensional, unitless "slope correction factor S" that subtracts -1 from the slope of the log-log linear model slope ($b_{BWIB}$) of $\log_{10}$BWIB *vs* $\log_{10}(M)$: $S = (b_{BWIB} - 1) / b_{BWIB}$

Now, we can transform the BWIB data into S-normalized Biomass (fnB):

$\log_{10}(fnB_i) = S * \log_{10}(BWIB_i)$

If we do not need raw $fnB_i$ values, we may do everything in log-space, since $10^{(\log10znB)}$ will produce very large values (leading to numeric overflow in extant computers).

Alternatively, the "S-normalized Biomass" SnB, may be numerically reduced for convenience (giving birth to the "S-normalized, reduced Biomass", SnrB, with much lower numerical values), while preserving the size spectrum pattern, as: $\log_{10}(SnrB) = \log_{10}(SnB) - \max(\log_{10}(SnB))$,

and then: $SnrB = 10^{\log10(SnrB)}$

After S-normalizing and reducing, rescaling is needed, by using the S-scaled correction factor S' ("S-prime"), where $S' = \log BWIB_{total} / SnrB_{total}$

The $znrB_i$ data are then back-transformed into the scale of B, into the "backtransformed" normalized biomass (bNB) : $bNB_i = SnrB_i * S'$

This avoids the linearizing effects of multiplying BWIB mass by 1/mass, which may produce linearizing spurious artifacts on a log-log-scale. Indeed, solution S3 produced spectra with higher variability, which proves that some data may be better represented though S3 (direct S-normalization), than when dividing by a log-scaled vector (vectors "k" or "w" in solutions S1 and S2).

An important decision, when conducting the final rescaling step, using the ratio of biomass sums (e.g., $D = BWIB_{total} / NBSS_{total}$ ), is whether to use the global biomass across the whole data set (the complete biomass of the whole sample, including linear, non-linear, and border sections), or the total biomass within the useful linear section only, that is actually used for slope and intercept calculations. Using only the useful linear section for biomass sum calculations avoids considerable uncertainty, especially regarding the intercept of the linear model.

Solution S3 seems, at first sight, to be the best, cleanest and most elegant solution. Yet, multiplying BWIB by any factor enhances (corrects) the negative slope sections, but also augments the positive sections of peaks and domes, leading to a hyper-distorted shape, especially when the spectrum contains extensive positive slope sections (Suppl. Mat., Figs SM 5a, 5b). Solution S3 seems to provide potentially useful results only for perfectly linear data (or at least, when there are no sections with positive slope in the final spectrum).

Simply put, Solutions S1 and S2 rotate the original bin-with-inflated spectrum (making it steeper by -1), while solution S3 stretches (distorts) it vertically, which also makes the positive sections steeper. This leads to a hyper-distorted biomass spectrum shape in S3 (i.e., an overstretched representation of the variability). Thus, the S3 approach cannot be generally recommended to represent complex natural size spectra (Suppl. Mat., Figs SM 5a, SM 5b).

### *3.7.4 Comparing binned methods with KDE*

The comparison of solution S1, S2, and S3 bNBS and bNNS (backtransformed normalized numbers spectrum, i.e. the abundance-body-mass spectrum) with each other, with linear binning, with kernel density estimation (KDE), and with the original input data (input power law "lambda" values), showed that all three solutions seem to represent well the original size spectrum slope (negligible bias in slope values, under low-variability scenarios).

The comparison of variability represented by the different calculation methods (S2, S2, S3) proved much more complex. While S1 and S2 give identical results, S3 produces plots with conspicuously higher variability , often with exotic, unexpected patterns (Suppl. Mat., Figs SM 5a, SM 5b).

Kernel density estimation (KDE) is an excellent approach for the representation of the shape of distributions, such as abundance-body-mass probability densities (total = 1), but it cannot provide any quantitative information on abundance or biomass (this is why we need bNNS and bNBS). The best way to check whether solutions S1, S2, and S3 are correct (representing most correctly the peaks and through of the original data), is by comparing with KDE (with different bandwidths, e.g. bw = 0.2, 0.25 and "SJ" algorithm), wich is only possible for bNNS (not for bNBS).

In all simulations, the comparison of bNNS obtained through solutions S1 and S2 with KDE showed a congruent shape, without any "exotic" unexplained features, artifacts, or hyper-distorted shapes. Conversely, the comparison of solution S3 bNNS with KDE (the un-normalized "gold" standard to portrait density distributions), with high-variability spectra (data that were dotted with discrete peaks and domes, on purpose) showed that S3 produces hyper-distorted spectra, especially in sections with positive slope, which was extremely pronounced in high-variability spectra (Fig Suppl mat 5c showing bNNS Solution 3 vs solutions 1 & 2 with KDE). Sections with slightly positive slopes in the bin-width inflated data became hyper-positive. This led to the observed hyper-distortion in S3, for data with positive sections (e.g., when there were peaks and domes in the spectrum).

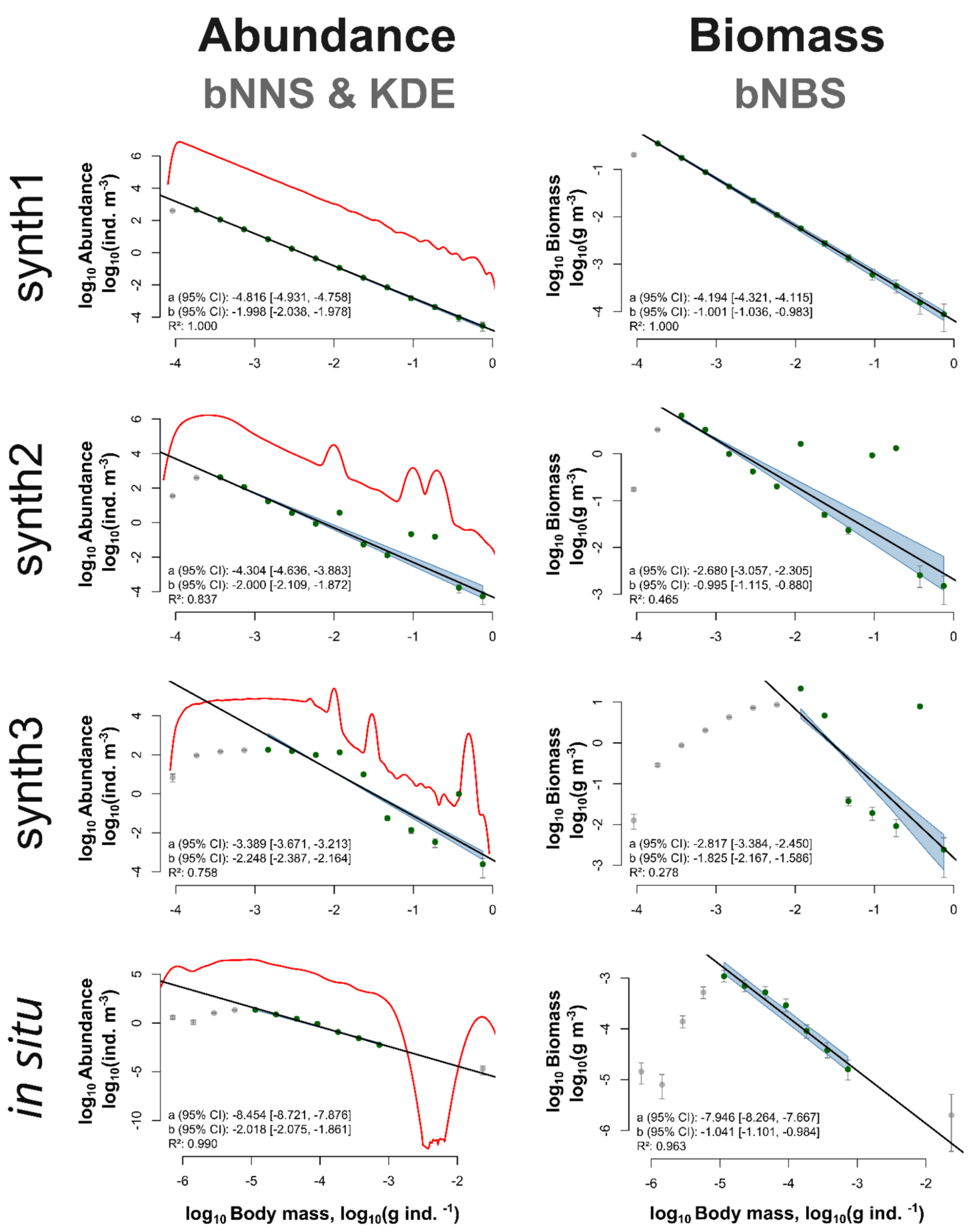

**Figure 6.** Examples of suggested analyses and plots (bNBS, bNNS, and KDE). bNBS: backtransformed normalized biomass spectrum. bNNS: backtransformed normalized numbers

spectrum. KDE: kernel density estimation (bandwidth selection: Silverman's 'rule of thumb', Silverman, 1986). syhth1: perfect power law distribution (near-zero variability); syhth2 : high-variability power law distribution data, with three peaks; syhth3: very high variability power law distribution data, with three sharp peaks; *in situ*: a tropical zooplankton sample, from Figueiredo et al. (2025). Vertical error bars: 95% bootstrap confidence intervals. Data for two-step bootstrapped robust linear regression (blue 95% confidence envelopes) were selected from the maximum to the last non-zero bin. Green dots: bins used for linear model fit.

The simulations and tests listed above show that approaches S1 and S2 can be recommended for the calculation of backtransformed spectra of abundance and biomass (bNNS and bNBS). Thus, a multi-step bootstrap routine that accounts for the original sample size (and all sources of uncertainty in the steps leading to the fitted robust regression linear model) was developed and successfully implemented for bNNS (with KDE for the unbinned visualization of peaks and throughs) and bNBS (using solution S1 and direct binning), and tested with numerous synthetic and *in situ* datasets (Fig. 6, Suppl., Mat. Fig. SM 6a,b).

# 4. Discussion

## *4.1 Estimation of biomass richness of ecosystems ($D_o$, $NBSS_o$, , $B_o$, and $bNB_o$) and the maximum carrying-capacity constant (kappa)*

Quantitative estimation of biomass richness (i.e., the mean vertical position of the spectrum) in ecosystems, from binned spectra (e.g., NBSS or bNBS), is far from trivial. First, it is important to recognize that biomass is never size-independent. Any reported biomass value $B_i$ (e.g., 0.3 g $m^{-3}$) is always a sum of individual body-mass values over a specific, discrete body-mass interval “i” (Fig. 6). In other words, it is impossible to determine a quantitative biomass value $B_i$ without explicitly defining the size range “$w_i$” over which it is measured, including its lower and upper bounds and its characteristic midpoint $M_i$ (the geometric mean of the body size interval “i” in linear space, which corresponds to the euclidean midpoint in log space).

The fundamental quantity of interest, that is independent of the arbitrarily chosen size range (i.e., independent of the chosen bin width in a histogram), is the biomass density $D_B$ (biomass per unit body mass, e.g., g $m^{-3}$ $g^{-1}$) (*sensu* Kerr & Dickie, 2001). The most parsimonious and correct description of an ecosystem is thus the “biomass density *vs* body mass spectrum”, i.e. its continuous “biomass density function” (biomass per unit body mass):

$$D_{Bi} (M_i) = D_{Bo} * M_i^{beta} \qquad (1)$$

Here, $D_B(M)$ denotes biomass density per unit body mass (Fig. 6). It is independent of the chosen body-mass interval width w, but varies as a function of body mass $M_i$ and of the ecosystem-specific “density scaling parameter” $D_{Bo}$. The exponent beta is often approximately beta = - 1 in pelagic ecosystems (Table 1, Schwamborn, 2025). The density scaling parameter $D_{Bo}$ (i.e., the log-equivalent linearly scaled intercept) is defined as the biomass density $D_B$ at log(M) = 0 (i.e., M = 1 in the chosen mass unit, such as 1 g carbon, or 1 cubic millimeter in biovolume-based spectra). The exact numerical value of $D_{Bo}$ depends on the selected units of body mass (e.g., g, mg, $\mu m^3$) and biomass. The units of $D_B$ and DB0 are typically expressed as $m^{-3}$, analogous to NBSS units. The ecosystem-specific biomass-density scaling parameter $DB_0$ can be assessed from the intercept of a log-log linear regression (e.g., for a linear model with base-10 logarithms: $DB_0=10^{intercept}$).

When binning is conducted with regularly spaced linear bins (constant bin width w, Fig. 7c and 6d), log–log projected spectra of biomass density $D_B(M)$ and binned biomass $B_i(M)$, have exactly the same slope "beta" but differ in their intercepts (the relationship between these two intercepts is simply the log of the harmonic mean bin width w': $\log B_o = \log D_{Bo} + \log w'$).

The integral of the linear biomass density function $D_B(M)$ above gives the cumulative biomass-body mass function ($C_B$ *vs* M), which has an exponent of beta +1 (except in the exact beta = −1 case, when $C_B(M) \sim \ln M$, which is not a power-law):

$$C_{Bi}(M_i) = (D_{Bo} / (beta+1)) * M^{beta+1} \qquad (2)$$

When beta = -1, equation 2 formally yields $\beta+1 = 0$, and thus, at first sight, a $C_B$ spectrum exponent of zero (perfectly flat spectrum), but this is not possible for a cumulative distribution (cumulative functions always monotonously increase with M, as long as $D_B(M) > 0$), so in the special case of beta = -1, the exponent is undefined, and the $C_B$ vs M function gives a simple logarithmic, non-power-law spectrum (for beta = -1 : $C_B = \sim \ln(M)$ ). This may be related to the "nearly flat ecosystem" portrayed by some authors (e.g., Boudreau and Dickie, 1992, Maxwell & Jennings, 2006, Trebilco et al., 2015), and may have been erroneously confounded with the supposed existence of a nearly flat linear (non-cumulative) B *vs* M relationship. Most importantly, binning is a finite-interval aggregation (or discrete approximation), not a continuous integral (as is the cumulative biomass function), which may have been confounded by earlier studies that wrote about integrals in this context.

The fact that for $\beta = -1$, the cumulative $B_c$ *vs* M exponent is undefined (because the cumulative function then is no longer a power law, but a logarithmic function), may also help explain why MLE (Edwards et al., 2017), uses the cumulative function of the abundance spectrum (with exponent ≈ −2) for plotting purposes. In contrast to binning-based approaches (e.g., NBSS and bNBS), cumulative (e.g., the "log cumulative distribution", "LCD" method, Bolker, 2008) and MLE methods thus are not suited to evaluate the biomass spectrum (with slope and intercept) and to visualize its shape (i.e., its mean slope, height, bumps, peaks, gaps, and domes), and to visually compare the particular biomass values of specific size classes.

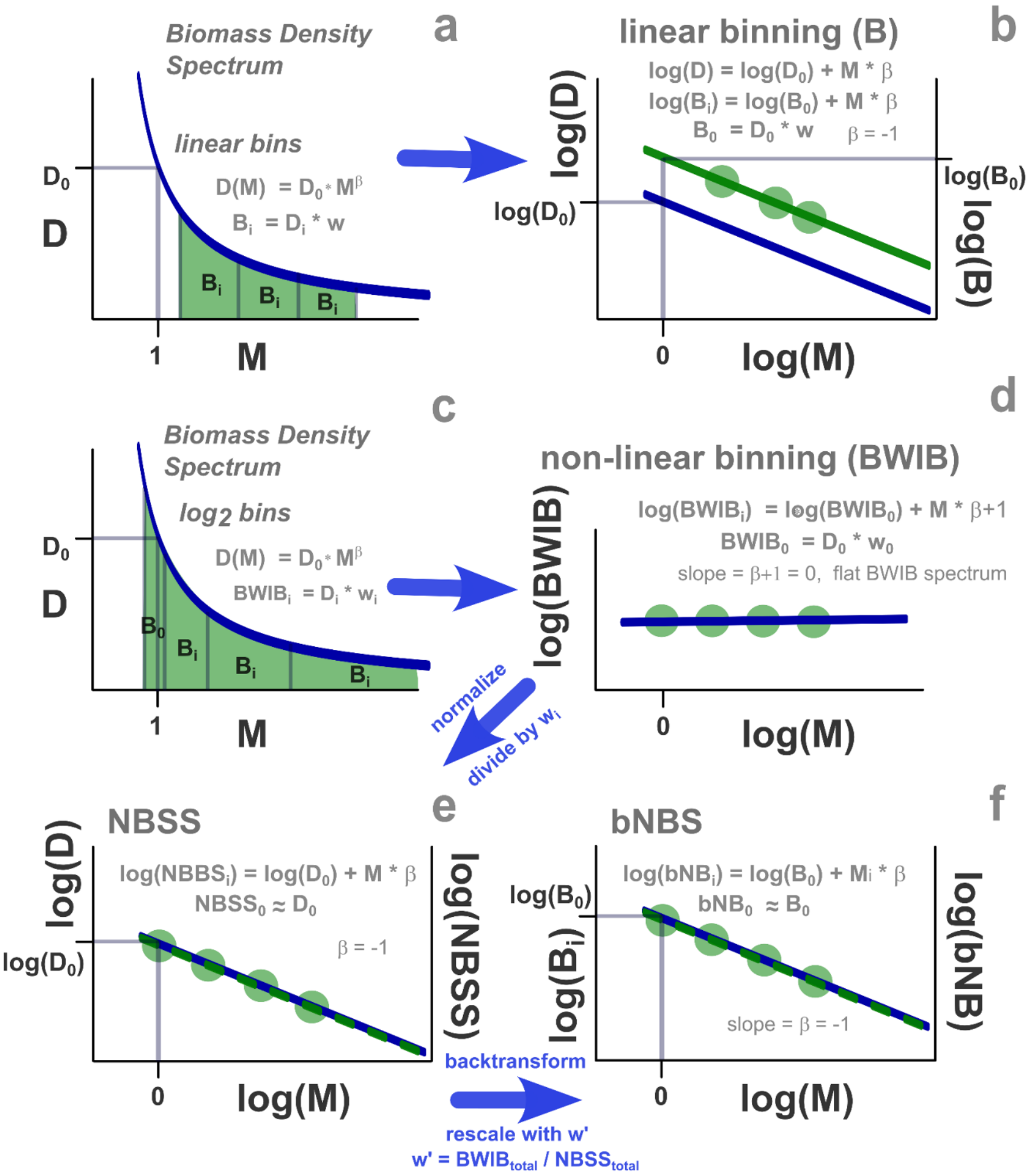

**Figure 7. From Body Mass (M) data to the bNBS.** Schematic representation of power-law distributions of biomass obtained from natural ecosystems and their relevant exponents (log-linear slopes) and scalings (log-linear intercepts). D(M): the biomass density distribution function $B_i$: binned Biomass, which are discrete-bin (linear binning) integrals over $D_B$. BWIB: bin-width inflated biomass (the BWIB spectrum slope is distorted by non-linear binning). w': effective mean bin width, or "biomass-weighted harmonic mean bin width "( w' = $BWIB_{total}$ / $NBB_{total}$)

The correct formulation for binned biomass Bi, expressed per unit sampled volume (e.g., per cubic meter), across the body-mass interval "i", is a function of the geometric-mean body mass $M_i$ and the body-mass interval width $w_i$.

By the way, the equation "$D_o = B_o / w_o$" also applies to the flat BWIB vs body mass spectrum ($D_o = BWIB_o / w_o$), which may be a useful way to estimate the uncertainty in $D_o$ estimates from non-linear binning vectors. The exact numerical value of $NBSS_o$ depends on the chosen units of M (e.g. g, milligram, cubic micrometers, etc.), the chosen value of $M_o$, log(M) = 0 (e.g M = 1 g), and the units of B. Most importantly, $NBSS_o$ is independent of bin width w.

Although $NBSS_o$ and $\mathbf{D_{Bo}}$ are approximately identical ($NBSS_o \approx D_{Bo}$) the distinction is mainly how they are used and interpreted, considering that NBSS is a finite-bin approximation of the continuous $D_B(M)$ function. Below are short descriptions of $D_{Bo}$, $NBSS_o$ and $B_o$, to outline how they are interrelated:

- **$D_{Bo}$: "biomass density scaling parameter" of the continuous biomass-density spectrum $D_B(M_i)$ *vs* $M_i$.** $D_{Bo}$ depends on space and mass units used , but is independent of bin width.
- **$NBSS_o$:"NBSS scaling parameter", derived from the intercept of the NBSS linear model** (e.g., for a **NBSS log-linear model** with base-10 logarithms: $NBSS_o = 10^{intercept}$). $NBSS_o$ may be considered a numerical approximation (a proxy) of $D_{Bo}$. Similarly to $D_{Bo}$, the exact value of $NBSS_o$ also depends on space and mass units used, but is independent of bin width.
- **$B_o$: "binned biomass scaling parameter", derived from the intercept of the bNBS linear model**. $B_o$ (or "$bNB_o$") depends on space and mass units used, and on the binning vector used. For a **bNBS log-linear model** with base-10 logarithms: $B_o=10^{intercept}$,

- **$bNB_o$:"bNBS scaling parameter", derived from the intercept of the bNBS linear model** (e.g., for a **bNBS log-linear model** with base-10 logarithms: $bNB_o=10^{intercept}$). $bNB_o$ may be considered a numerical approximation (a proxy) of $B_o$. Depends on space and mass units used, and on the binning vector used to calculate the biomass-weighted harmonic average mean bin width w'.

where: $bNB_o = NBSS_o * w' \approx D_{Bo} * w' \approx B_o$ (3)

There has been a lot of confusion regarding $NBSS_i$, biomass density $D_{Bi}$, and binned biomass $B_i$ being integrals or derivatives of each other. Some authors treated NBSS as if it was equivalent to NBSS = dB/dM, but the correct equation is actually $NBSS_i = B_i / d M_i$, which is not a derivative. Actually, from the above, it becomes clear that $NBSS_i$ is a representation of mean biomass density $D_{Bi}$ within bin "i": $NBSS_i = BWIB_i / w_i = D_{Bi}$.

It is important to keep in mind that each $BWIB_i$ (and thus, for each $bNBS_i$) is the area of of $D_{Bi} * w_i$ over the discrete size interval "i", while $NBSS_i$ is the mean biomass density in the size interval "i". While there is a consensus that the biomass *vs* body mass log-linear slope in pelagic ecosystems is approximately beta = -1, there is still no overall consensus regarding the exact values of $NBSS_o$ and $D_o$. These can hardly be universal constants, since the numerical value of $D_o$ depends on the total ecosystem biomass, the position of $M_o$ (the position of M where log(M) becomes zero, i.e., on the units of M, which define the value of M = 1, whether M is in "g" or "milligram"), and specifically of the biomass density of the size class "i" with $M_i$ = 1 g. Clearly, there is a need for an international standard binning vector (e.g. $\log_2$-scaled breaks) and a universal standard reference mass $M_o$ (e.g., 1 g). This is urgent and essential for any evaluations regarding worldwide size patterns of $D_o$ and $NBSS_o$ (e.g., in comparisons of worldwide quantitative biomass levels, across regions, communities, sampling methods, and time periods).

For binned biomass spectra (for a given bin width $w_i$),

$$NBSS_i = BWIB_i / w_i \approx D_B(M_i) = D_o * M_i^{beta} \quad (4)$$

$$\log(NBSS_i) = alpha + \log(M_i) * beta \quad (5)$$

Tentatively based on the above equations, we may state that the binned biomass intercept estimate $B_o$ may be assessed from $NBSS_o$ , the NBSS intercept alpha:

$$B_o = = bNB_o = NBSS_o * w' = 10^{alpha} * w' \quad (6)$$

**Table 2:** Scaling exponents of common size spectrum equations and their typical values in pelagic ecosystems.

| Quantity | Scaling exponent y(M) ~ M^( … ) | Typical values of the exponent |
|---|---|---|
| | | |
| Power-law numbers vs body mass exponent, N(M) | - lambda | -2 |
| Abundance density $D_A(M) = dC_A/dM$ | - lambda | -2 |
| Cumulative Biomass $C_A(M)$ function (integral of $D_A(M)$) | - lambda + 1 | -1 |
| Biomass density $D_B(M) = dC_B/dM$ | β = - lambda + 1 | -1 |
| Cumulative Biomass $C_B(M)$ function (integral of $D_B(M)$) | β + 1<br>(except when beta = -1) | undefined for beta = -1 |
| Biomass per linear bin (constant w) | β | -1 |
| Biomass per log2 bin, not normalized = BWIB “flat spectrum” | β + 1 | 0 |
| Biomass per log2 bin, normalized = NBSS | β | -1 |
| Biomass per log2 bin, backtransformed and normalized = bNBS | β | -1 |
| Abundance per log2 bin, normalized = NNSS | β -1 | -2 |
| Abundance per log2 bin, backtransformed and normalized = bNNS | β -1 | -2 |

Considering the above, the biomass at the intercept of the bNBS ($bNB_o$) may then be described as:

$$\log(bNB_i) = \log(bNB_o) + \log(M_i) * beta \qquad (7)$$

where

$$bNB_o = NBSS_o * w' \qquad (8)$$

NBSS is thus a representation of biomass density (more precisely: the finite-bin average of the continuous biomass density distribution $D_B$), while the bNBS is a representation of binned biomass B.

If the $\log(NBSS_i)$ vs $\log(M_i)$ slope is beta = -1, then the underlying $\log(B_i)$ vs $\log(M_i)$ slope is also beta = -1, and the underlying log(D) vs log(M) relationship is also beta = -1, all of which are different from the log (dB/dM) vs (M) slope of beta-1.

For example, several authors ( e.g., Boudreau and Dickie,1992, Kerr & Dickie, 2001 Benoit & Rochet, 2004) observed NBSS log-linear slopes of b = approx. -1 in various aquatic ecosystems, and concluded that the log biomass density $D_B$ *vs* log body mass M slope would be close to zero (i.e., equal biomass densities in all sizes, not a power law) , based on the erroneous assumption that the linear-scale $D_B$ *vs* M function is a derivative of the linear-scale B *vs* M function.

The confusion comes from mixing $D_B$, B, and NBSS, i.e., a continuous density function ($D_B$), and a finite-bin approximation of $D_B$ (the NBSS), which is not the same as a derivative, neither a finite-bin derivative. NBSS is therefore not a derivative of biomass B, nor a "finite-bin derivative" of B. It is a finite-bin average density function, obtained by normalizing binned biomass B by the bin width w. Or, simply put, the NBSS is an *empirical approximation* of the underlying density function.

Simply put, NBSS is a bin-normalized biomass quantity ($NBSS_i = BWIB_i / w_i$), therefore NBSS ≠ dB/dM. $NBSS_i$ may indeed be described as the mean density $D_{Bi}$ in each bin, but not as the derivative of B (i.e., dB/dM). The confusion may arise because $D_B$ (M) can indeed be described as the derivative of the cumulative biomass function $C_B$ (M), but $D_B$ (M) is not the derivative of binned biomass B *vs* body mass M (where cumulative biomass $C_{Bi}$ is the accumulated total biomass from $M_o$ up to $M_i$).

Several previous authors (e.g. Platt and Denman, 1978, Blanco et al., 1994) , stated that if B = M * N, then B(w)/w = N(w), and erroneously concluded the NBSS may indeed represent Abundance. Yet the term M*N(w) / w is only equal to N (w) in the special case where w = M, which cannot be correct across a size spectrum (w = M may be correct at one specific size bin, in the best case). Simply put, the erroneous idea of NBSS being representative of abundance is probably due to a confusion of dividing by mean size and dividing by size range.

Now we can easily estimate $B_o$, i.e. the biomass level at "$M_o$", where $M_i$ = 1 (i.e., the bin width at log(M) = 0 ), from the bNBS:

$D_{Bo} = bNB_o / w'$ (9)

Finally, the fact that the function of $D_B$ ($m^{-3}$) vs M (g) is the first derivative of the cumulative biomass $C_B$ (g $m^{-3}$) vs M (g) function explains its correct units and dimensions, and the very nature of the $D_B$ ($m^{-3}$) vs M (g) spectrum. Simply put, $D_B$ ($m^{-3}$) vs M (g) describes the change in biomass accumulation per unit body mass.

Within PETS ("predator-prey-efficiency theory of size spectra") theory (Schwamborn, 2025), the universal carrying-capacity parameter *kappa* , is conceptually and numerically similar to $D_o$ and $NBSS_o$. PETS describes, among other processes, hypotheses and mechanisms, how top-down processes may constrain biomass below a universal carrying-capacity spectrum across all body-mass classes. If there is indeed a theoretical maximum biomass value for each size class in marine pelagic ecosystems, *kappa* may be simply called the mass-specific **maximum carrying-capacity constant** (Schwamborn, 2025).

Within the PETS framework, top-down mechanisms keep the biomass in each size class below the maximum carrying-capacity spectrum $NBSS_{max}$, thereby stabilizing its NBSS log-linear *intercept* $alpha_{max}$ *, which is simply* $alpha_{max}$ = *log(kappa).* Simply put, *kappa* can be described as the $NBSS_{o,max}$ value (i.e., the y-intercept) of the $NBSS_{max}$ representing the maximum carrying capacity spectrum, for example, estimated by fitting a linear regression to the maximum values (i.e., the 95th percentiles, an approximation of the maximum) of NBSS values within each body-mass bin (Schwamborn et al., submitted).

$NBSS_0$, $DB_0$, and kappa are independent of bin width, whereas $B_0$ (expressed in conventional biomass units, e.g., g $m^{-3}$) depends linearly on the chosen reference bin vector. Although kappa, $NBSS_0$, and $D_{B0}$ may indeed represent a relevant ecological scaling parameters, they possesses non-intuitive and dimensionally complex units. In contrast, $B_0$ corresponds to an easily interpretable biomass quantity in intuitive, simple units.

*4.2 Normalized biomass is still biomass*

Currently, there is still considerable confusion and ambiguity regarding what NBSS values actually represent (e.g., Sato et al., 2015, Sprules and Barth, 2016). Since it was demonstrated here, through extensive numerical simulations, that NBSS accurately represents biomass density (not biomass flux or abundance), the next logical step is a clear definition of the relevant variables, each with its unique variable name, for B and for dB/dM. Thus, to distinguish the biomass variable B from dB/dM, a new, consistent terminology is proposed:

We may call the back-transformed biomass values simply “biomass” (B), or “back-transformed normalized biomass” (bNB). Accordingly, the plot and model (i.e., the biomass *vs* body mass relationship) may be called simply “biomass-body-mass-spectrum”, or bNBS (“back-transformed normalized biomass spectrum”).

Within the context of NBSS "theory," a frequently stated misconception is that the intercept of the linear NBSS model represents a proxy of primary production or "ecosystem productivity", as if the intercept were necessarily located at the small-size end of the spectrum (e.g., Zhou, 2006; Marcolin et al., 2013; Sato et al., 2015). In reality, the intercept can occur anywhere along the size spectrum, depending on the range of body sizes included in the analysis and the reference body mass value $M_0$. For example, the intercept may correspond to the upper end of the sampled size range, particularly when most sampled individuals have body masses substantially smaller than $M_0 = 1$ g, which means that the intercept is located at the upper limit of the dataset (e.g., Fock et al. 2026; Schwamborn et al., submitted). When body mass is expressed in grams, the intercept corresponds to the predicted $NBSS_0$ value at a body mass of 1 g, regardless of whether 1 g falls within, below, or above the sampled size range (e.g., Fock et al. 2026; Schwamborn et al., submitted).

### *4.3 The good old NBSS - not rebuked, but improved*

Many critical evaluations regarding spurious autocorrelation bias in commonly used methods (e.g., Schwamborn, 2018), have led to the necessity of rebuking and replacing them by newly developed methods (e.g., Schwamborn, et al., 2019). Conversely, in this study, the commonly used NBSS method is preserved and confirmed in its principles, although several pitfalls, misconceptions, and shortcomings have been detected and addressed. The newly proposed bNBS method, approach, and terminology intends to correct, complement, and improve the calculations, models, concepts and terminology in normalized binning-based methods. This addresses the urgent need for more consistent methods, models, and approaches, which is especially important in the context of increasingly available databases from semi-automatic imaging devices (Dugenne et al., 2024).

Some authors have tried to solve these paradoxes by avoiding any normalization at all, by fitting linear models directly to non-normalized, nonlinearly ($\log_2$) binned biomass-body mass BWIB data (e.g., Maxwell & Jennings, 2006; Trebilco et al., 2015). Although this approach may appear reasonable at first glance, using non-normalized BWIB often produces flat ($b = 0$) or positive ($b > 0$) biomass–body mass log-log-linear model slopes (Maxwell & Jennings, 2006; Trebilco et al., 2015). Such patterns represent a distorted portrayal of natural ecosystems and may lead to erroneous and misleading interpretations - specifically, the impression that all size classes contain equal total biomass, or even that larger organisms have greater biomass than smaller ones. These distorted patterns in BWIB do not reflect ecological reality but instead arise as artifacts of non-normalized non-linear binning. Therefore, I would advocate for not using the BWIB as a representation of natural ecosystems (but rather, only as an intermediate calculation step).

Edwards et al. (2017) compared several binning–based methods with their non-binned maximum likelihood estimation (MLE) of the size spectrum slope “b”. Although in their simulations, MLE proved superior to many (mostly exotic and unpopular) non-normalized binning–based methods, it seems that it was not superior to the common normalized binning approach (e.g., NNSS). When examining the results of Edwards et al., 2017, it becomes clear that their normalized binned method (NNSS, called “LBNbiom” in Edwards et al., 2017, 2020) actually produces exactly the same results (near-zero bias in the estimation of the power-law numbers exponent lambda = 2) as their proposed (but much more complex, and less intuitive) non-binned MLE method. Thus, our study is in agreement with the results of Edwards et al. (2017) in confirming the correctness and reliability of normalized binned methods.

Yet, in a subsequent study, Edwards et al. (2020) showed that binning underestimates the uncertainty in the sampling and binning process (as in Schwamborn et al., 2019), which is indeed an important argument for the non-binned MLE method. The uncertainty in the original size data is obviously underrepresented in binned (simplified) data. The apparent difference in results (slope "b" value) obtained by MLE and the normalized binned method (NNSS sensu Vandromme et al., 2012 = LBNbiom sensu Edwards et al., 2017, 2020) in the Edwards et al. (2020) study, may be due to the peculiarities of the fish size datasets used in their study. More likely, this observation by Edwards et al., 2020, may be due to an error in applying binning-based methods (considering that in their Fig. 1, their "LBbiom" and "LBNbiom" plots show exactly the same numerical results, which can only explained by an error). Furthermore, their recommended MLE method has several disadvantages over normalized binned methods. For example, MLE is not able to estimate the intercept of the size spectrum linear model (biomass at M = 1 g ind.-1), and does not provide any intuitive linear-shaped graphs, and has therefore not become very popular among marine ecologists (Ersoy et al., 2025).

The original *in situ* biomass density ($D_B$) cannot be observed, described or sampled directly (any sampling effort produces a discretely binned $B_i$ estimate, not the continuous $D_B(M)$ function). Yet, the $D_B(M)$ function can be assessed indirectly through binning methods (e.g., by the NBSS plot method), or by using smooth kernel estimators of the probability density function, or by numerical differentiation (i.e. obtaining the first derivative) of the cumulative $C_B(M)$ function (Fig. 5). One main advantage of the NBSS method is that it provides a simple and intuitive way to visualize the $D_B(M)$ function of any given ecosystem.

In contrast to the $D_B(M)$ function, the cumulative $C_B(M)$ function can indeed be obtained directly from *in situ* observations. For example, $C_B(M_i)$ can be estimated using cumulative distribution approaches such as the LCD ("log cumulative distribution") method (Vidondo et al., 1997; Bolker, 2008; Rogers, Blanchard & Mumby, 2014, Edwards et al., 2017). Alternatively, maximum-likelihood estimation (MLE, Edwards et al., 2017) can be used to estimate the underlying body-mass scaling (Edwards et al., 2017, 2020), from which the $C_B(M)$ function can be easily plotted *a posteriori*, although the MLE approach is not explicitly based on fitting the cumulative function.

Yet, since the $C_B$ *vs* M plot is always necessarily monotonically increasing (independently of the size spectrum shape), cumulative plots are much less informative (and evidently non-intuitive) for visualizing, analysing, and interpreting the shape of the underlying *in situ* ecosystem size spectrum. Binning (NBSS and bNBS) and kernel (KDE) methods are evidently the only suitable choice to obtain linear model intercepts, and to visually compare the biomass of specific size classes, and to identify the position and magnitude of relevant troughs, domes, gaps, and peaks in the spectrum, for the whole ecosystem, and for key taxa (Schwamborn et al., 2025).

The combination of quantitative binning (bNNS and bNBS) and non-parametric kernel density estimation (KDE) proposed here, intends to address the importance of intuitive, simple plotting methods and the relevance of avoiding binning artifacts and oversimplifications, while providing quantitative estimates of many key partners such as abundance, biomass, and size spectra slopes and intercepts.

*4.4 A plea for the use of a standardized binning vector, and a standardized NBSS -to-bNBS conversion factor $w_R$*

The proposed bNBS method (i.e., normalization, redimensionalization, and rescaling) is a big improvement over the presently used NBSS. Further relevant improvements are  the equations for estimating $B_0$, and the the conceptually novel carrying capacity parameter kappa (*sensu* Schwamborn, 2025), and the robust multi-step bootstrap, which is a statistical improvement over common binned linear regression, considering the propagation of uncertainty throughout sampling and analysis.

The dimensions, units and totals are correct and intercomparable, and the y values a and the intercept value of a w-normalized spectrum such as the NBSS and NNSS, are independent of the binning vector used.

Yet, the absolute biomass that is seen in non-normalized, de-normalized or backtransformed biomass plots (i.e., the “y” values and the intercept value, in a bNBS plot) cannot be compared, if different bin widths are used. Although the totals are quantitatively identical within the same data set, different binning schemes will always produce different "y" values (as in any linearly binned histogram, where simply, y values are proportional to the global bin width w).

Most importantly, the effective numerical value of w′ is specific for each particular dataset (and the binning vector used). Consequently, w′ cannot be used as a universal conversion factor from NBSS to bNBS across datasets and ecosystems. Rather w′ is a dataset-specific biomass-weighted harmonic mean bin width, that can be used for the back-transformation of NBSS to bNBS within a given dataset, when the objective is to recover biomass values corresponding to the original binning scheme, and to compare biomass level within a given database (e.g., to compare biomass B between specific size classes, taxa or assemblages, within a given ecosystem).

For quantitative comparisons of biomass B levels across ecosystems, regions, or time periods, an internationally defined standard reference bin width ($w_R$) is needed (i.e., a *standardized NBSS-to-bNBS conversion factor* $w_R$) . The reference bin width ($w_R$) would provide a common biomass scale for expressing the back-transformed spectrum, independently of the dataset-specific effective harmonic mean bin width (w′). For example, $w_R$ could be defined as $w_R = 1$ g, provided that the corresponding mass unit (1 g) is explicitly adopted. Such a convention would make bNBS-derived biomass levels, and particularly parameters such as $B_o$, quantitatively comparable across datasets, while w' would remain the appropriate dataset-specific factor for recovering biomass on the original binning scale.

If $w_R$ is simply defined as 1 g and is used to convert NBSS to bNBS, then the numerical values of current spectra based on this unit (g carbon biomass, as in Fock et al, 2026) do not change at all (each quantitative $bNB_i$ value , e.g., in g $m^{-3}$ would then simply be $NBSS_i$ multiplied by 1). Since linear binning (Suppl. Mat., Fig. SM1a), which is standard for single-species (e.g., fish growth studies, e.g. Schwamborn et al., 2023) and single-assembly studies, is not feasible for datasets spanning many orders of magnitude (needed to describe any meaningful ecosystem), non-linearly binned biomass (i.e., BWIB) is the only available biomass representation, which has to be transformed and interpreted in a meaningful way (w-normalized into NBSS and then linearly back-transformed into bNBS). Cross-checking the correctness of an ecosystem-wide bNBS locally, for selected assemblages, by comparing it with linear-binned histograms (for slopes and intercepts, overall shape, peaks, domes and gaps, testing several different bin widths) seems to be a useful approach for cross-checking, but keep in mind that any biomass estimate (and thus the intercept "a", of the bNBS) is always a function of bin size (w). Our main proposal here must be to explicitly define a reference-bin $w_R$ and specifically, a unity reference $w_R$ =1, under which common NBSS values become numerically invariant representations of quantitative binned biomass B, without the need of a an explicit numerical resampling procedure.

Since there is no need to explicitly resample the continuous NBSS function (i.e., to take each value from the NBSS data and model predictions and multiply it by 1), this means that one may simply use the previously published common NBSS plots and simply rename the y axis from NBSS (g $m^{-3}$ $g^{-1}$) to quantitative binned biomass B (g $m^{-3}$) if the procedure of converting from NBSS to bNBS with the unity reference bin with (wR = 1 mass or volume unit) is explicitly defined in the materials and methods, and ideally also in all figure legends.

Currently, there is no standard for a universal binning vector, nor for reference bin width $w_R$, and unfortunately, most NBSS studies use unique units and vectors (specifically adapted for each dataset). For example, one could imagine that the most simple, intuitive $log_2$-binning vector (breaks = ... , 0.25 g, 0.5 g, 1 g, 2 g, 4 g, ... , with right-closed binning) used herein and in numerous previous studies could be universally defined as the standard binning vector. This simple, very common $log_2$ vector, has, however the disadvantage, that it is not fully centered at the intercept, i.e., the log(M) = 0 value is not located at the center (i.e., at the geometric mean) of any bin (i.e., M = 1 g lies at a bin boundary, not at the geometric center of any bin).

A solution to this shortcoming could be a "$log_2$-unity" binning vector, consisting of a fixed number of bins (e.g., 80 bins) whose midpoints follow a strict log2 progression, and where the central bin ($i_0$) is exactly centered at $M_0$ = 1. Bin boundaries in the "$log_2$-unity vector" are defined such that each midpoint is the geometric mean of its lower and upper break. And the "$log_2$-unity" vector is indeed log2-scaled (i.e., "octave-scaled" with breaks = ..., 0.176 g, 0.353 g, 0.707 g, 1.414 g, 2.828 g, 5.656 g, ... ).

Another potential candidate could be a vector with $w_0$ = 1 g, and $M_0$ = 1 g as its (geometric mean) midpoint, but such a "double unity vector" means that it cannot be $log_2$-scaled (it has to be the $log_{2.618}$-scaled, i.e., exactly "1+golden ratio"-scaled, i.e. breaks = ..., 0.0901 g, 0.2361 , 0.6181 g, 1.6181 g, 4.2361 g, ... ). Having a binning vector where the $i_0$ segment has log(M) = 0, and bin width $w_0$ = 1 g, has a certain advantage, since then all the intercepts $NBSS_0$, $bNB_0$, $D_{B0}$, and $B_0$ become equal. They are indeed only equal in the special case where the reference bin width is $w_R$ = $w_0$ = 1, while in all other cases, $D_0 \neq B_0$.

Another quite obvious, well-known and often stated urgent necessity, is the need for standardized sampling methods (e.g., nets and mesh sizes), and units of mass (often used units: g C, mg organic C, mt wet weight, g dry weight, mg dry weight, $microns^3$, $mm^3$, etc.) and space ($km^2$, $m^2$, $L^{-1}$, $m^3$) in quantitative ecosystem research, which are currently not standardized across methods, communities, and ecosystems, hampering the inclusion of key ecosystem properties and size-specific processes, in global climate models.

Choosing large-scale space units (area, e.g. $km^2$ for regional and global scale models) and a large reference mass unit and standard reference bin width $w_R$ (e.g., 1 mt, as in hundreds of Ecopath models) may be conceptually preferable to using small-scale units (e.g., cubic meters per litre, as in Dungenne et al., 2024), since small-sized organisms (e.g. phytoplankton) may comfortably fit into large space and mass units, whereas large organisms (e.g., fish and mammals) can hardly be represented in small (micron-sized) mass and spatial units.

Most conveniently and interestingly, multiplying by wR = 1 mass or volume unit (e.g. 1 g, 1 mg $ind.^{-1}$, 1 $mm^3$, or 1 cubic micrometer per individual), means that all previously published NBSS studies do not have to be numerically converted to obtain quantitative biomass B. Based on rationale, all previous NBSS can be simply and directly re-interpreted as quantitative biomass. For example, instead of stating that in the Fock et al. (2024, 2026) dataset, that was based on g C units, NBSS represents a “proxy of biomass” or a “representation of biomass density”, actually each single and NBSS value represents the exact quantitative value of binned biomass B, in g C $m^{-3}$, considering a *a posteriori* linear binning (back-transformation) with a linear reference bin width $w_R$ of 1 g.

The emerging possibility to convert NBSS to biomass B without changing basically anything in plots, intercepts, slopes, and models, may also lead us to a new look at NBSS. Instead of a merely "proxy of biomass" or as a "histogram approximation of biomass density" only, we may also interpret NBSS as quantitative biomass (B) estimate, under the assumption of a unity vector (i.e., as if the NBSS was explicitly resampled using a linear, binning vector with a constant bin with of 1g, and each $NBSS_i$ value was multiplied by 1). Thus each $NBSS_i$ value simply represents a $B_i$ estimate, with 1 g bin width.

Whatever the chosen universal units of mass (e.g., 1 g carbon, 1 mg wet mass, 1 mt wet weight, or 1 $mm^3$ biovolume) and space (e.g, $m^3$, litres, or $km^2$), and thus, of the reference bin width $w_R$ (for conversion from $D_B$ to B units, or from NBSS to bNBS units) will look like in the near future, standardization of units is extremely urgent and important to allow quantitative intercomparisons across regions, seasons, and years, e.g., for modeling efforts and time series studies in the context of ongoing global climate change. One of the most important contributions of this study may be the insight that, under this rationale, previously published NBSS values can be reinterpreted and utilized, without changing any numerical values, as quantitative biomass (B) values corresponding to a standardized reference bin width of 1 mass unit, thereby facilitating their use in quantitative modeling efforts.

*4.5 Looking at intercepts and slopes of biovolume-based NBSS databases*

Biovolume-based models (total biovolume vs individual biovolume, Vandromme et al., 2012) are conceptually and dimensionally similar to biomass-body-mass models, since biovolume is often approximately proportional to biomass. Biovolume NBSS models have become increasingly common in plankton ecology, due to the simple geometric transformations from images to 3D volumes (e.g., ellipsoids or spheres, for thousands of individuals), that can be semi-automatically obtained from modern imaging devices. Thus, many recent zooplankton NBSS publications are in biovolume units (e.g., Vandromme et al., 2012, Lira et al., 2024, Dugenne et al., 2024, Figueiredo et al., 2025). Yet, we must consider that large-sized gelatinous organisms have high biovolume-carbon ratios, thus biovolume models may be considerably flatter than biomass models (Schwamborn et al., submitted). Thus, for standardization, I suggest that whenever possible, pelagic ecosystem size spectra be presented in units of carbon biomass (converted from biovolume to carbon mass by taxon-specific conversion factors), which is where we may expect a slope of approximately b = -1.

In our simulations, we observed a perfect linear correlation between intercept (a) and slope (b). Similarly, Dugenne et al. (2024) reported an almost perfect correlation between a and b estimates in their dataset. In their biovolume-based NBSS , the intercept corresponded to 1 cubic micrometer (i.e., log(x) = 0) representing the smallest size class (the size of picoprokaryotes, the smallest known phytoplankton). Their reference value  log(x) = 0 was located at the lower extreme of the biovolume vector rather than near its midpoint. Consequently, their intercept was strongly influenced by slope variation. Thus, the apparent invariance of the NBSS intercept across polar (biomass—rich) and tropical (oligotrophic) waters reported in their study may reflect relatively high picoplankton biomass in tropical waters (associated with the lower extreme positioning of the reference size class), i.e., an approximate (log-scale) invariance of picoplankton biomass, rather than an evidence of a globally invariant pelagic ecosystem biomass.

*4.6 The NBSS slope "b": biomass flux or mass-conversion efficiency?*

The simulations and tests conducted in this study confirmed the concept that power-law distributions of biomass density, biomass, NBSS, and bNBS, all have an identical log-linear slope of beta = -1 (Fig. 6). The methodological rationale illustrated in Figure 6 also explains well the origin of the non-biomass units in NBSS. Simply put, NBSS has non-biomass units because it represents mean binned biomass density (not mean binned biomass), as had been recognized by several previous authors (Blanco et al., 1994). Yet this is the first study to verify the precision and accuracy of binned methods and to numerically confirm the validity of the normalized biomass approach. Most importantly, this study rebukes an earlier, rather simplistic simulation study (Prothero, 1986), that implied that the NBSS slope is due to spurious artifacts, not a representation of actual ecosystem structure. Thus, this study is in agreement with the extensive analyses of Blanco et al. (1994), who also criticized the conclusions of Prothero (1986), but did not yet prove a simulation-based validation of the NBSS approach.

The existence of a power-law distribution in any physical or biological system means ubiquity of scale invariance, and the absence of a dominant scale (Kadanoff, 1966, Wilson, 1971). Scale invariance in natural ecosystems means that the same fundamental processes govern at all size classes (Schwamborn, 2025). The absence of a dominant scale in power laws (as opposed to a normal or lognormal distribution) means that there is no dominant size class in the ecosystem, or dominant taxon. The apparent "dominant taxon" (e.g., copepods in marine mesozooplankton, or myctophids in marine fish) is merely a function of the arbitrarily chosen body size range and sampling method.

Rather than a "NBSS theory" (*sensu* Platt & Denman, 1978, Blanco et al., 1994, Marcolin, 2013, Marcolin, et al., 2013, Hernández-Moresino et al., 2017) the rationale discussed here and by previous authors (e.g., Blanco et al., 1994) can be considered a practical, methodological framework for binning-based biomass studies, as opposed to a general theory of ecosystem size spectra (such as PETS, "predator-prey-efficiency theory of size spectra", Schwamborn, 2025) that is a theoretical explanation of the processes that shape size distributions in natural ecosystems. Ecosystem size spectrum theory and methodological considerations should always be clearly distinguished, yet they remain closely interrelated and have important implications for one another.

Most importantly, the exact value of beta = -1 of the biomass-body-size log-log-linear slope, confirmed herein, has profound implications for our understanding of the functioning of ecosystems (Schwamborn, 2025). If we consider that body mass and trophic level are stringently correlated within a size-structured food web, the biomass-body-mass-relationship dB/dM may be a relevant index of how biomass flows through the spectrum, of variations in predator-prey mass ratio PPMR, and mass-specific, tropic-level (TL) - specific efficiency "E", where E = d(log(M)) / dTL, (Schwamborn, 2025).

A plethora of quantitative sampling studies (e.g., Dugenne et al., 2024) have demonstrated that for pelagic marine ecosystems, there is a ubiquitous d (logB) / d log(M) slope of approximately b = 1. The ubiquitous constancy of b = - 1 means that there is a universally constant, proportional, weight–scaled, weight-specific biomass transfer efficiency: (dB / dM) dM-1. This observation implies different aspects of the equilibrium and proportionality (Schwamborn, 2025): it means that that B and M are inversely proportional ( B ~ 1/M ), that log(M) and log(B) are linearly (negatively) proportional, and finally, and that "E" is also linearly (negatively) proportional to "log(PPMR)" (Schwamborn, 2025).

Thus, the recent PETS theory ("predator-prey-efficiency theory of size spectra", Schwamborn, 2025) explains the ubiquitous b = 1 slope, generally observed in pelagic ecosystems as:

b = E / log(PPMR)

PETS (Schwamborn, 2025) also contains numerous detailed descriptions of compensation and equilibrium mechanisms to explain the b = 1 slope, e.g., based on top-down trophic cascades, resource-limitation stress, and "size spectra-specific optimal foraging theory" (SOFT, Schwamborn, 2025).

Yet, the relationship between biomass and body mass (i.e., the slope of the double logarithmically transformed biomass spectrum), should not be considered an index of flux, as the terms "flux" and "rate" imply units of time (speed of change per unit time or mass transfer, or energy transfer per unit time). The size spectrum, whether in units of abundance or biomass, does not contain any dimensions that are stringently related to time (although some authors have related size to turnover time, e.g., Platt and Denman, 1978). Instead, the slope of the biomass-body-mass spectrum may rather be regarded as a measure of size-specific efficiency, or, more precisely, of mass-specific efficiency (if there is a size-structured food web, as shown in Figueiredo et al., 2020).

The decrease in biomass with trophic level ( d log(B) / d TL ) is explicitly defined as the mass specific trophic efficiency "E" (Schwamborn, 2025). Analogously, the decline of biomass with body mass ( b = d log(B) / d log(M) ) can also be regarded as a form of trophic efficiency. That means that, within PETS (Schwamborn, 2025), when the size difference between predator and prey is very large, the system is less efficient, within a stringent proportionality. If we consider the dTL / dB relationship to be the mass-, biomass-, PPMR-, and TL-specific trophic efficiency, the slope "b" may be called the "biomass-body-mass-efficiency" BBME. If we concur with the available data for pelagic ecosystems, that show a general ubiquity of approximately b = -1, this would mean that in our planet's pelagic ecosystems, we have a generally constant BBME trophic efficiency (i.e., a universal proportionality of B and 1/M), that is scale-invariant and independent of taxonomic group, temperature, and metabolism (Schwamborn, 2025). This study demonstrated that the ubiquitous beta = - 1 is not an artifact of model construction and linearizing artifacts, but rather, a key property of our planet's pelagic ecosystems.

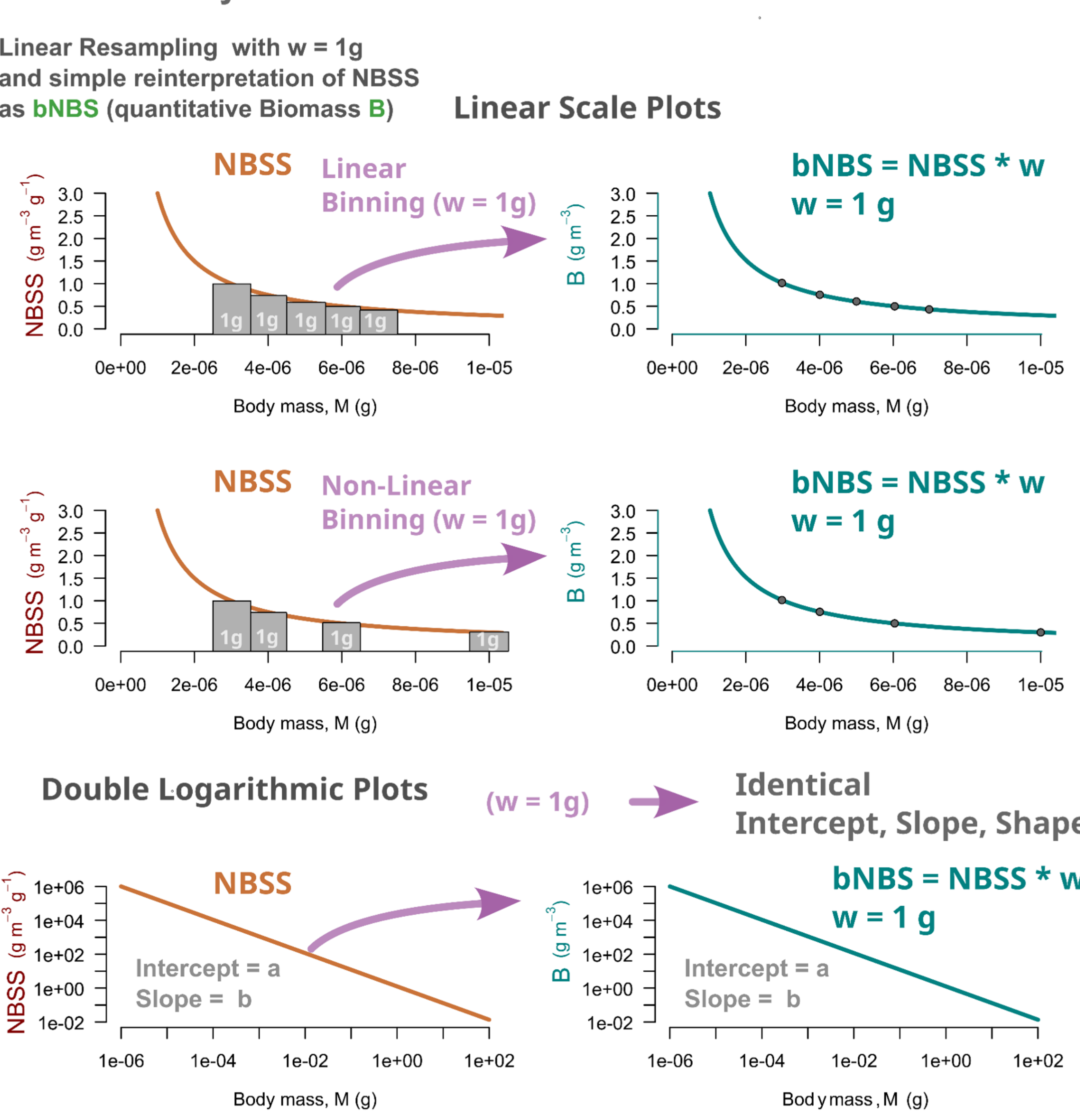

**Fig. 9 Simple transformation from NBSS to bBNS using the unity bin width (numerical value = 1 mass unit).** To obtain bNBS, each NBSSi value is multiplied by the unity bin width value (w = 1 g). The corresponding areas (grey bars) give the $bNBS_i$ value for each $NBSS_i$, which is numerically identical (note the identical "y" axis values for NBSS and bNBS), but has different y units. Above: linear binning. Center: Non-linear binning (usually used for building NBSS). Below: Double logarithmic (log-log) plots.

*4.9 Towards a new SIZE framework*

Considering the above, the term "NBSS theory" seems misleading because it implies that the calculation procedures of NBSS constitute, in themselves, a meaningful biological theory. This term has also led to a conflation of two fundamentally different contexts and levels of analysis: ecological theory, which seeks to explain processes (e.g. trophic interactions, allometric relationships, and physiological constraints, that may generate the ubiquitous b = -1 NBSS slope), and methodological considerations (i.e., the interpretation of the NBSS units).

In reality, the NBSS just represents one particular statistical and graphical formulation used to visualize size-dependent abundance and biomass distributions (it's a method, not a specific theory). Yet, NBSS parameters are generated through a sequence of choices involving sampling, body-size measurement, binning, normalization, scaling, and statistical fitting, and therefore cannot be interpreted independently of these procedures.

Statistical and epistemological considerations must be taken into account in many other methods used to analyze size spectra (e.g. NNSS, MLE, LCD, etc.), not only in NBSS plots. I therefore propose the broader term "*Size-spectrum Inference, Scaling, and Estimation*" (SIZE) theory, for considerations within this broader methodological and epistemological context. Unlike the term "NBSS theory", SIZE theory is not restricted to bin-width-normalized methods (e.g., NBSS) and does not imply any specific parameter (e.g., biomass) or normalization procedure. The SIZE framework may thus encompass analyses based on abundance, biomass, density, cumulative distributions, and other representations of size- or mass-structured objects. Considering the ubiquity of power-law distributions, such methodological and epistemological considerations may also apply to many abiotic datasets (e.g., microplastics, detritus particles, and asteroids, e.g. Clauset et al., 2009; Lins-Silva et al., 2024). Considering the plethora of existing methods and approaches in various areas of research, one key objective of SIZE theory could be to develop a consistent set of epistemological concepts, definitions, statistical methods and intuitive visualization techniques for size-spectrum analysis throughout quantitative science.

From an *Erkenntnistheorie* perspective, several quantities, such as density distributions $D_x(M)$, cannot be directly observed; they are inferred from datasets produced by particular sampling, detection, resolution, binning, normalization, and estimation procedures. This distinction between natural objects, their observation and possible mathematical representations is the very foundation of epistemology of measurement (Heisenberg, 1958; Duhem, 1914; Miyake, 2013) and should be central to the development of any relevant and robust SIZE theory.

This paper, based on the work of Blanco et al. (1994) and many others (e.g. Platt & Denman, 1978, Blanco et al., 1994, Marcolin, 2013, Marcolin, et al., 2013, Hernández-Moresino et al., 2017), may be another step forwards into the development of a "Size-spectrum Inference, Scaling, and Estimation" (SIZE) theory.

In the context of this paper, the most important SIZE considerations were the rejection of the "NBSS-is-biomass-flux" hypothesis (NBSS represents the biomass density spectrum, not biomass flux), the distinction between density and cumulative distributions and between integrals and binned approximations. The insight of the need to treat BWIB as a separate, well-defined quantity, came from the analysis of the BWIB *vs* M distribution as being intrinsically distinct from the body-mass-biomass (B *vs* M) spectrum, which led to the insight of dimensional invariance of BWIB ( BWIB has exactly the same units and physical dimensions as B, but is a distorted representation of B). These definitions and insights were key to understanding the meaning of the units of the NBSS (i.e., to the solution of the "NBSS-Biomass-Abundance-paradox") and led to the path towards the development of the novel bNBS.

*4.8 Recommendations and conclusions*

The results of this study and the rationale detailed above lead us to a short list of ten basic recommendations for size spectra research:

I.) Use meaningful, correct units in all plots and equations; if you use the procedure of converting from NBSS to bnBNS with a unity reference bin with (e.g. $w_R$ = 1 g) explicitly define this (and all units of each step) in the materials and methods, and ideally also in the figure legends;

;

II.) Inform whether direct or indirect binning of biomass was used;

III.) Whenever possible, use direct binning of biomass (not B = A * M);

IV.) Provide plots of abundance (KDE, bNNS), and biomass (bNBS) spectra, always indicating the section used for linear models of abundance and biomass (e.g., Figueiredo et al., 2025);

V.) To investigate the shape of the spectrum (peaks and troughs), test different binning vectors (bNNS and bNBS), and bandwidths (KDE), and different types of size spectra plots and transformations (e.g., Schwamborn et al., 2025);

VI.) For quantitative comparisons of linear models (intercepts of bNBS and bNNS linear models), use a standardized $\log_2$ binning vector (e.g., breaks = ... , 0.25, 0.5, 1, 2, 4, ... , right-closed);

VII.) Avoid naïve analyses and interpretations of variations in linear model slope and intercept within the same analysis, better: use a slope-independent estimator of biomass (e.g., total predicted B, mean B, B at a selected size range), instead of reporting only the intercept (which is often correlated to the slope);

VIII.) Conduct quantitative analysis only within the most data-rich section of the spectrum (avoid sections with border effects, empty bins, and sampling selectivity effects);

IX.) For quantitative comparisons of abundance and biomass at specific sizes (biomass-at-size), provide the exact size range (or body mass range) used.

X.) Prefer robust regression (RR) over OLSR (common "ordinary least squares regression") to fit linear models (less prone to outlier effects, see Figs. Suppl. Mat. SM 5 a,b), with permutation-based "p" values and two-step bootstrap 95% confidence intervals.

This study confirms the accuracy and reliability of normalized binned methods in determining the ecosystem-wide size spectrum slope, the existence of scale-invariant processes (as in any power-law distribution), and the "universal biomass-body-mass inverse proportionality" (slope b = -1) paradigm for pelagic ecosystems (Schwamborn, 2025). Yet, we detected several important issues (such as inadequate units, terminology, and linearizing artifacts), and provided suggestions for improvements in this important and rapidly expanding area of research, based on large databases and semi-automatic methods.

This has obvious implications for the analysis of size spectra of living organisms, microplastics and other particles in natural ecosystems (Lins-Silva et al. 2024). Also, these results may be relevant for many other areas of research, that investigate the shape of power-law distributions (due to the ubiquity of scale invariance, the absence of a dominant scale, and heavy tails), such as Economy, Finance, Linguistics, Sociology, Geography, Epidemiology, Geophysics, Astrophysics, Computer Science, and Information Theory (e.g., Pareto, 1897, Zipf, 1949, Newman, 2005, Barabási & Albert, 1999, Axtell, 2001, Pastor-Satorras & Vespignani, 2001, Clauset et al., 2009, Gabaix, 2009, Piantadosi, 2014).

Considering the oversimplification and linearization by binned methods, natural ecosystems probably exhibit a much larger variability than suggested in most size spectra databases and large-scale studies (Fock et al. 2024, Dugenne et al., 2024). Investigating this variability due to local, regional and global processes that affect natural size spectra (e.g., Lira et al., 2024, Figueiredo et al., 2025) is an important field of ecosystem research, which may benefit from the proposed improvements and suggestions.

This study highlights that for any quantitative study (not only for size spectra), providing the exact body mass range, when informing abundance or biomass values, while considering only the most data-rich section of the spectrum, should be standard. Yet, in most quantitative ecological and ecosystem models, only the sampling method (i.e., mesh size), is provided. Sampled size amplitude, mean size, and size spectrum slope are generally ignored in ecosystem models, except in modern size-spectra-based individual models (Dalaut et al., 2025). Far from being a merely formalistic detail, this simple conclusion (the impossibility of determining general, size-independent biomass or abundance estimates, a "dominant" taxon, or a dominant size class, in a power-law distribution, where mean size, median size, and dominant size are a function of the arbitrarily chosen size range) has far-reaching consequences for the future of quantitative ecology and for our ability to predict the future of our planet's ecosystems.

## Acknowledgments

Many thanks to Gaby Gorsky and Marc Picheral for introducing me to the ZooScan method more than two decades ago, which opened new paths towards the analysis of zooplankton size spectra. Many thanks to all students, postdocs, and colleagues in UFPE's zooplankton lab for companionship, unity, friendship, and intensive work on zooplankton size spectra with ZooScan, especially to Lúcia Gusmão *(in memoriam)* and Sigrid Neumann Leitão. Many thanks to Gordon "Gordie" Swartzman for introducing me to the R software, language and environment more than 20 years ago. Many thanks to Nathália Lins-Silva and Catarina Marcolin, for co-developing our common zooplankton size spectrum analysis research stream. Special thanks to Catarina Marcolin for discussions about the "theory of NBSS", many years ago, which further encouraged me to dedicate myself to the problems and paradoxes of this method. Many thanks to M. L. "Deng" Palomares for encouraging me and for pushing me deeper into the work on size-based methods, more than 17 years ago. Many thanks to Daniel Pauly for inspiring me, along many years (since 1991), to conduct in-depth theoretical work on this subject. The author is especially indebted to the TRIATLAS size spectra working group for attending many size spectra workshops and numerous inspiring discussions. Many thanks especially to Emilio Marañon, Cristina González-García, Henrike Andersen, Mathilde Dugenne, Lars Stemmann, Tim Dudek, and Heino Fock, for important discussions during the TRIATLAS size spectra workshops. To Dodji Soviadan for in-depth discussions on UVP size spectra. Many thanks to Arnaud Bertrand for inspiring comments, continuous support, and friendship. To Denise F. M. C. Schwamborn and Wesley de Oliveira Neves for revising an early version of the text. Many thanks to Simone A. Lira, Gabriela Figueiredo, Claudeilton de Santana, Eduardo T. Paes, and Iurick Saraiva Costa for important comments. Special thanks also to Gleice Souza Santos for pushing me to prioritize this paper over many other important paper projects. Many thanks to Lorenzo Fant for inspiring, controversial discussions and important insights regarding the dimensional invariance of BWIB, during the recent AtlantEco workshop in Ponta Delgada.

## Artificial Intelligence (AI) declaration

During the preparation of this manuscript, the author used artificial intelligence (AI) tools (Grammarly, Google Scholar, and ChatGPT) for text editing (e.g., grammar and spelling corrections) and literature searches. No AI tools were used to generate ideas, scientific content, text structure, figures, equations, analyses, or interpretations. All content is the original work of the author.

# References

Atkinson, A., Rossberg, A. G., Gaedke, U., Sprules, G., Heneghan, R. F., Batziakas, S., ... , Frangoulis, C., 2024. Steeper size spectra with decreasing phytoplankton biomass indicate strong trophic amplification and future fish declines. Nature Communications, 15 (1), 381.

Auerswald, K., Wittmer, M.H.O.M., Zazzo, A., Schaufele, R., Schnyder, H., 2010. Biases in the analysis of stable isotope discrimination in food webs. J Appl Ecol 47:936–941

Axtell, R. L. (2001). Zipf distribution of U.S. firm sizes. Science, 293(5536), 1818–1820.

Barabási, A.-L., & Albert, R. (1999). Emergence of scaling in random networks. Science, 286(5439), 509–512.

Bensen, M.A., 1965. Spurious correlations in hydraulics and hydrology. J Hydraul Div Proc Am Soc Civ Eng 91:35–42

Benoit, E., Rochet, M.J. (2004) A continuous model of biomass size spectra governed by predation and the effects of fishing. J. Theor. Biol. 226, 9–21 ()

Blanco, J.M., Echevarría, F., & García, C. M., 1994. Dealing with size spectra: some conceptual and mathematical problems. Scientia Marina, 58, 17–29.

Brett, M.T., 2004. When is a correlation between non-independent variables “spurious”? Oikos 105:647–656. https://doi.org/10.1111/j.0030-1299.2004.12777.x

Chayes, F., 1949. On ratio correlation in petrography. J Geol 57:239–254

Clauset, A., Shalizi, C. R., & Newman, M. E. J. (2009). Power-law distributions in empirical data. SIAM Review, 51(4), 661–703.

Cuesta, J. A., Delius, G. W., & Law, R. (2018). Sheldon spectrum and the plankton paradox: two sides of the same coin—a trait-based plankton size-spectrum model. Journal of mathematical biology, 76(1), 67-96.

Dalaut, L., Barrier, N., Lengaigne, M., Rault, J., Ariza, A., Belharet, M., Brunel, A., Schwamborn, R., Travassos-Tolotti M., Maury, O. (2025). Which processes structure global pelagic ecosystems and control their trophic functioning? Insights from the mechanistic model APECOSM. Progress in Oceanography, 235, 103480.

Dugenne, M., Corrales-Ugalde, M., Luo, J. Y., Kiko, R., O'Brien, T. D., Irisson, J. O., ... Vilain, M., 2024. First release of the Pelagic Size Structure database: global datasets of marine size spectra obtained from plankton imaging devices. Earth System Science Data, 16(6), 2971-2999.

Duhem, P. (1914). La théorie physique: Son objet et sa structure. Paris: Marcel Rivière.

Edwards, A. M., Robinson, J. P., Blanchard, J. L., Baum, J. K., Plank, M. J., 2020. Accounting for the bin structure of data removes bias when fitting size spectra. Marine Ecology Progress Series, 636, 19-33.

Edwards, A. M., Robinson, J. P., Plank, M. J., Baum, J. K., Blanchard, J. L., 2017. Testing and recommending methods for fitting size spectra to data. Methods in Ecology and Evolution, 8(1), 57-67.

Ersoy, Z., Evangelista, C., Larrañaga, A., Perkins, D. M., Sánchez-Hernández, J., Chonova, T., ... & Arranz, I. (2025). GLOSSAQUA: A global dataset of size spectra across aquatic ecosystems. Ecology, 106(3), e70050.

Figueiredo, G.G.A.A., Lira, S.M..A, Bertrand, A., Neumann-Leitão, S., Schwamborn, R., 2025. Zooplankton abundance and biovolume size-spectra in the western tropical Atlantic-From the shelf towards complex oceanic current systems. Marine Environmental Research, 204, 106906.

Figueiredo, G.G.A.A., Schwamborn, R., Bertrand, A., Munaron, J. M., Le Loc'h, F., 2020. Body size and stable isotope composition of zooplankton in the western tropical Atlantic. J. Mar. Syst. 212, 103449. https://doi.org/10.1016/j.jmarsys.2020.103449.

Fock, H., Andresen, H., Bertrand, A., Bitencourt, G., Carré, C., Couret, M., Díaz-Pérez, J., Dudeck, T., Dugenne, M., Duncan, S., Figueiredo, G. G. A., Frédou, T., Díaz, X. F. G., Ghebrehiwet, D. Y., González-García, C., Kiko, R., Landeira, J. M., Lira, S. M. A., Lüskow, F., Marañón, E., Melo, P. A. M. C., Neumann-Leitao, S., Eduardo, L. N., Stemmann, L., Schwamborn, R., 2024. Synthetic Pelagic

Biomass Size Spectra of the Tropical and Subtropical Atlantic - biovolume and carbon biomass data [Data set]. Zenodo. https://doi.org/10.5281/zenodo.13627093.

Fock, H., Andresen, H., Bertrand, A., Bitencourt, G., Carré, C., Couret, M., Díaz-Pérez, J., Dudeck, T., Dugenne, M., Duncan, S., Figueiredo, G.G.A., Frédou, T., Lucena-Frédou, F., Díaz, X. F. G., Ghebrehiwet, D. Y., González-García, C., Kiko, R., Landeira, J. M., Lira, S.M.A., Lüskow, F., Marañón, E., Melo, P. A. M. C., Neumann-Leitao, S., Eduardo, L. N., Stemmann, L., Schwamborn, R. (2026). Synthetic pelagic biomass size spectra of the tropical and subtropical Atlantic. Ecosphere, 2026;17:e70608. DOI: 10.1002/ecs2.70608

Freedman, D., & Diaconis, P. 1981. On the histogram as a density estimator: L 2 theory. Zeitschrift für Wahrscheinlichkeitstheorie und verwandte Gebiete, 57(4), 453-476.

Gabaix, X. (2009). Power laws in economics and finance. Annual Review of Economics, 1, 255–293.

Gómez-Canchong, P., Blanco, J. M., & Quiñones, R. A. (2013) On the use of biomass size spectra linear adjustments to design ecosystem indicators. Scientia Marina, 77(2): 257–268

Heisenberg, W. (1958). Physics and Philosophy: The Revolution in Modern Science. New York: Harper & Brothers.

Hernández-Moresino, R. D., Di Mauro, R., Crespi-Abril, A. C., Villanueva-Gomila, G. L., Compaire, J. C., Barón, P. J., 2017. Contrasting structural patterns of the mesozooplankton community result from the development of a frontal system in San José Gulf, Patagonia. Estuarine, Coastal and Shelf Science, 193, 1-11.

Huber, P. J. (1964) Robust Estimation of a Location Parameter. The Annals of Mathematical Statistics, 35, 73-101. https://doi.org/10.1214/aoms/1177703732

Kadanoff, L. P. (1966) Scaling Laws for Ising Models Near Tc. Physics Physique Fizika 2, 263–272 . DOI: 10.1103/PhysicsPhysiqueFizika.2.263

Kanaroglou P.S., 1996. On spurious correlation in geographical problems. Can Geogr 40:194–202

Kenney B.C., 1982. Beware of spurious self-correlations! Water Resour Res 18:1041–1048

Lins-Silva, N., Marcolin, C. R., Kessler, F., & Schwamborn, R. (2024). Influence of microplastics and biogenic particles on the size spectra of tropical estuarine and marine pelagic ecosystems. Science of the Total Environment, 927, 172244. https://doi.org/10.1016/j.scitotenv.2024.172244

Lira, S. M., Schwamborn, R., de Melo Júnior, M., Varona, H. L., Queiroz, S., Veleda, D., Silva, A. J., Neumann-Leitão, S., Araujo, M., Marcolin, C. R., 2024. Multiple island effects shape oceanographic processes and zooplankton size spectra off an oceanic archipelago in the Tropical Atlantic. Journal of Marine Systems, 242, 103942.

Marcolin, C.D.R. 2013. Plankton and particle biomass size spectra on the Southwest Atlantic: Case studies in tropical and subtropical areas (Doctoral dissertation, Universidade de São Paulo).

Marcolin, C.D.R., Schultes, S., Jackson, G.A., Lopes, R.M., 2013. Plankton and seston size spectra estimated by the LOPC and ZooScan in the Abrolhos Bank ecosystem (SE Atlantic). Continental Shelf Research, 70, 74-87.

Maxwell, T.A.D., Jennings, S., 2006. Predicting abundance-body size relationships in functional and taxonomic subsets of food webs. Oecologia, 150, 282– 290

Miyake, T. (2013). Underdetermination, black boxes, and measurement. Philosophy of Science, 80(5), 697–708. https://doi.org/10.1086/673729.

Mulder, C., Den Hollander, H. A., Hendriks, A. J., 2008. Aboveground herbivory shapes the biomass distribution and flux of soil invertebrates. PLoS One, 3(10), e3573.

Newman, M. E. J. (2003). The structure and function of complex networks. SIAM Review, 45(2), 167–256.

Newman, M. E. J. (2005). Power laws, Pareto distributions and Zipf's law. Contemporary Physics, 46(5), 323–351.

Pareto, V. (1897). Cours d'économie politique. Lausanne: F. Rouge.

Pastor-Satorras, R., & Vespignani, A. (2001). Epidemic spreading in scale-free networks. Physical Review Letters, 86(14), 3200–3203.

Pearson, K., 1897. On a form of spurious correlation which may arise when indices are used in the measurement of organs. Proc R Soc Lond 60:489–498

Piantadosi, S. T. (2014). Zipf's word frequency law in natural language. Psychonomic Bulletin & Review, 21, 1112–1130.

Platt, T., Denman, K., 1977. Organization in the Pelagic Ecosystem. Helgoländer Wissenschaftliche Meeresuntersuchungen 30 (1–4), 575–581.

Platt, T., Denman, K., 1978. The structure of pelagic ecosystems. Rapports et Procès-Verbaux des Réunions du Conseil International pour l'Exploration de la Mer, 173, 60–65.

Prothero, J. 1986. Methodological aspects of scaling in biology. Journal of Theoretical Biology, 118(3), 259-286.

Reed J.L., 1921. On the correlation between any two functions and its application to the general case of spurious correlation. J Wash Acad Sci 11:449–455

Rosenblatt, M., 1956. Remarks on some nonparametric estimates of a density and mode. Ann. Math. Statist, 27, 642-645.

Rossberg, A. G., Gaedke, U., Kratina. P., 2019. Dome patterns in pelagic size spectra reveal strong trophic cascades. Nature communications, 10(1), 4396.

Rousseeuw, P. J., & Leroy, A. M. (2003). Robust regression and outlier detection. John Wiley & Sons.

Sato, K., Matsuno, K., Arima, D., Abe, Y., & Yamaguchi, A. (2015). Spatial and temporal changes in zooplankton abundance, biovolume, and size spectra in the neighboring waters of Japan: analyses using an optical plankton counter. Zoological Studies, 54(1), 18.

Schwamborn, D. F. M. C. Marcolin C. R., Lins-Silva N., Brito-Lolaia M., Almeida A. O., Schwamborn, R., 2025. Size-niche specific processes between mero- and holoplankton stabilize the continuum in size spectra of tropical estuarine and marine ecosystems. Journal of Marine Systems, 104152. https://doi.org/10.1016/j.jmarsys.2025.104152.

Schwamborn, R. 2025. Towards a compleat theory of ecosystem size spectra. arXiv preprint, arXiv:2509.00023. https://arxiv.org/abs/2509.00023

Schwamborn, R., 2018. How reliable are the Powell-Wetherall plot method and the maximum-length approach? Implications for length-based studies of growth and mortality. Rev. Fish Biol. Fish. https://doi.org/10.1007/s11160-018-9519-0.

Schwamborn, R., Dudeck, T., Carré, C., Bittencourt Farias, G., Couret, M., Díaz Pérez, J., Dugenne, M., Guerra Araújo Abrantes Figueiredo, G., Frédou, T., Lucena-Frédou, F., Díaz, X. F. G., Yemane, D., González-García, C., Kiko, R., Hernández-León, S., Landeira, J. M., Lira, S. M. A., Lüskow, F., Dalaut, L., Maury, O., Marañón, E., Mendes de Castro Melo, P. A., Neumann-Leitão, S., Müller, M. N., Ekau, W., Nole Eduardo, L., Stemmann, L., A. Bertrand (submitted). Temperature, trophic state, and carrying capacity govern pelagic size spectra in the Atlantic Ocean (submitted)

Schwamborn, R., Mildenberger, T. K., Taylor, M. H., 2019. Assessing sources of uncertainty in length-based estimates of body growth in populations of fishes and macroinvertebrates with bootstrapped ELEFAN. Ecol. Model. 393, 37–51.

Sheather, S. J. and Jones, M. C. (1991). A reliable data-based bandwidth selection method for kernel density estimation. Journal of the Royal Statistical Society Series B, 53, 683–690. doi:10.1111/j.2517-6161.1991.tb01857.x.

Sheldon, R. W., A. Prakash, Sutcliffe Jr., W. H., 1972. The Size Distribution of Particles in the Ocean. Limnology and Oceanography 17 (3), 327–340.

Sheldon, R.W., Sutcliffe Jr, W.H., Paranjape. M.A., 1977. Structure of pelagic food chain and relationship between plankton and fish production. Journal of the Fisheries Board of Canada, 34(12), 2344-2353.

Sheldon, R. W., Sutcliffe Jr., W. H., 1973. Size of Particles, Sinking, and Trophic Structure in the Ocean. Limnology and Oceanography 18 (2), 327–340.

Silverman, B. W. 2018. Density estimation for statistics and data analysis. Routledge.

Sprules, W. G., Barth, L. E. 2016. Surfing the biomass size spectrum: some remarks on history, theory, and application. Canadian Journal of Fisheries and Aquatic Sciences, 73, 477–495.

Thompson, G. A., Dinofrio, E. O., & Alder, V. A. (2013). Structure, abundance and biomass size spectra of copepods and other zooplankton communities in upper waters of the Southwestern Atlantic Ocean during summer. Journal of Plankton Research, 35(3), 610-629.

Trebilco, R., Dulvy, N.K., Stewart,H., Salomon, A.K., 2015. The role of habitat complexity in shaping the size structure of a temperate reef fish community. Marine Ecology Progress Series, 532, 197–211.

Vandromme, P., Stemmann, L., Garcìa-Comas, C., Berline, L., Sun, X., Gorsky, G., 2012. Assessing biases in computing size spectra of automatically classified zooplankton from imaging systems: A case study with the ZooScan integrated system. Methods in Oceanography, 1, 3-21.

Venables W. N., Ripley B. D. (2002). Modern Applied Statistics with S, Fourth edition. Springer, New York. ISBN 0-387-95457-0, https://www.stats.ox.ac.uk/pub/MASS4/.

Wand, M. P. (1997). Data-based choice of histogram bin width. The American Statistician, 51(1), 59-64.

Wilson, K. G. “Renormalization Group and Critical Phenomena. I. Renormalization Group and the Kadanoff Scaling Picture” Physical Review B 4, 3174–3183 (1971). DOI: 10.1103/PhysRevB.4.3174

Zipf, G. K. (1949). Human Behavior and the Principle of Least Effort. Cambridge, MA: Addison-Wesley.

# Supplementary Materials

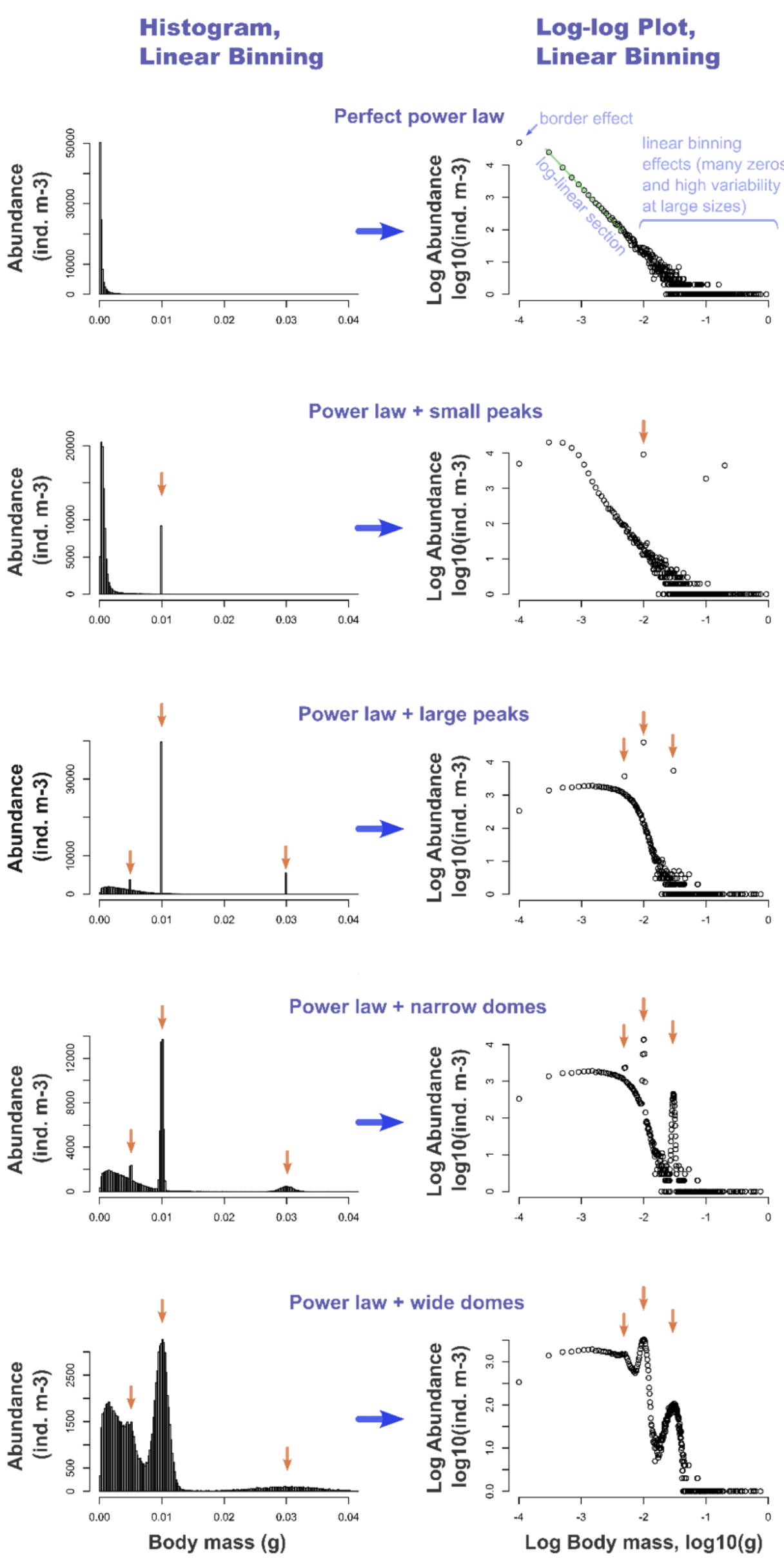

**Figure SM1a.** Examples of different types of datasets used in this study, plotted here with simple linear histograms (linear binning). Note the noisy and zero-rich sections for large-sized organisms (> -2 $\log_{10}$ (g)).

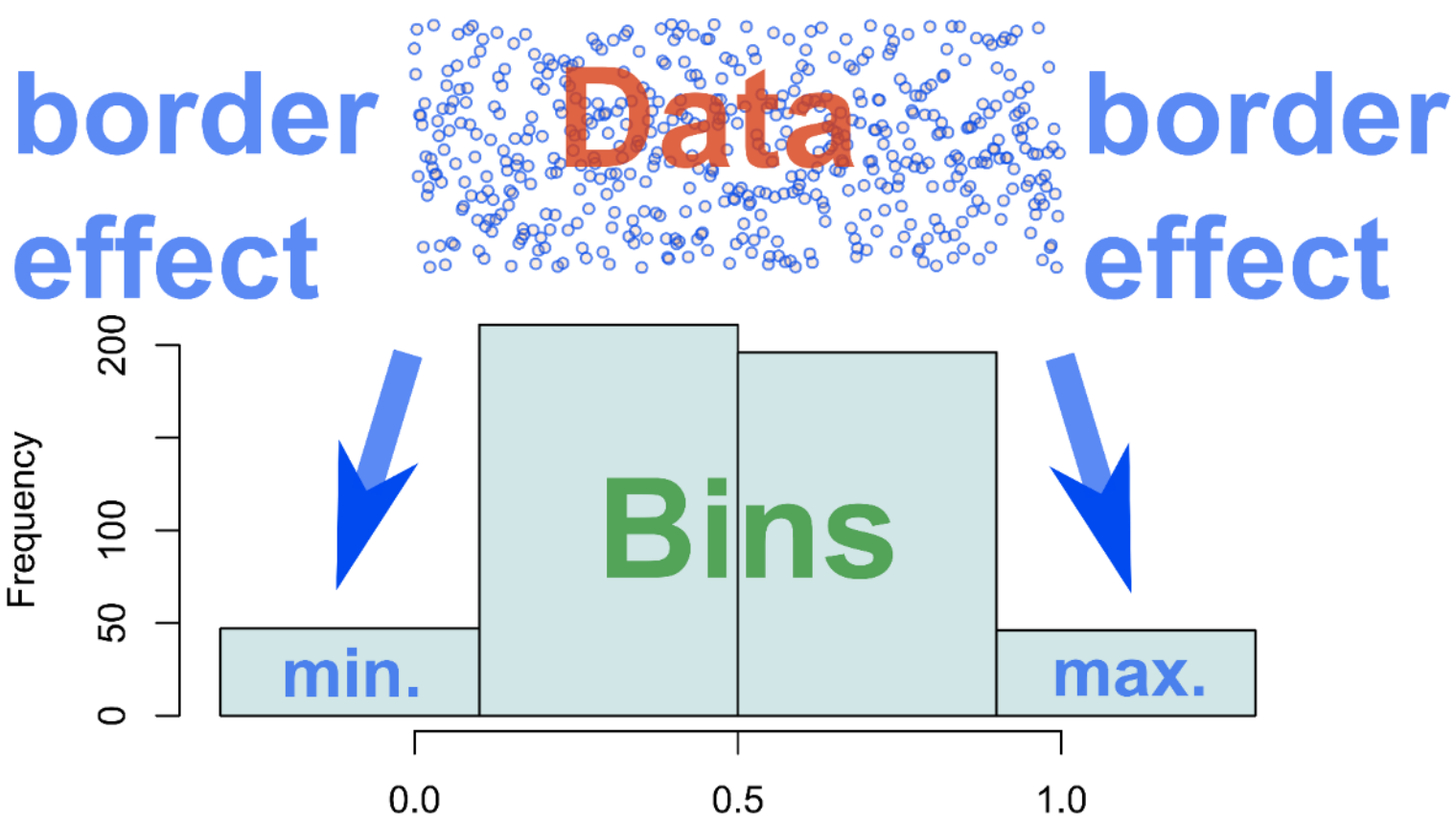

**Figure SM1b.** Border effects in a histogram. An example of original data vs binned data (histogram), depicting the uniformly distributed original data (above) and the border effects (below). Border effects always lead to underestimation of frequency and biomass at the lower and upper borders of histograms.

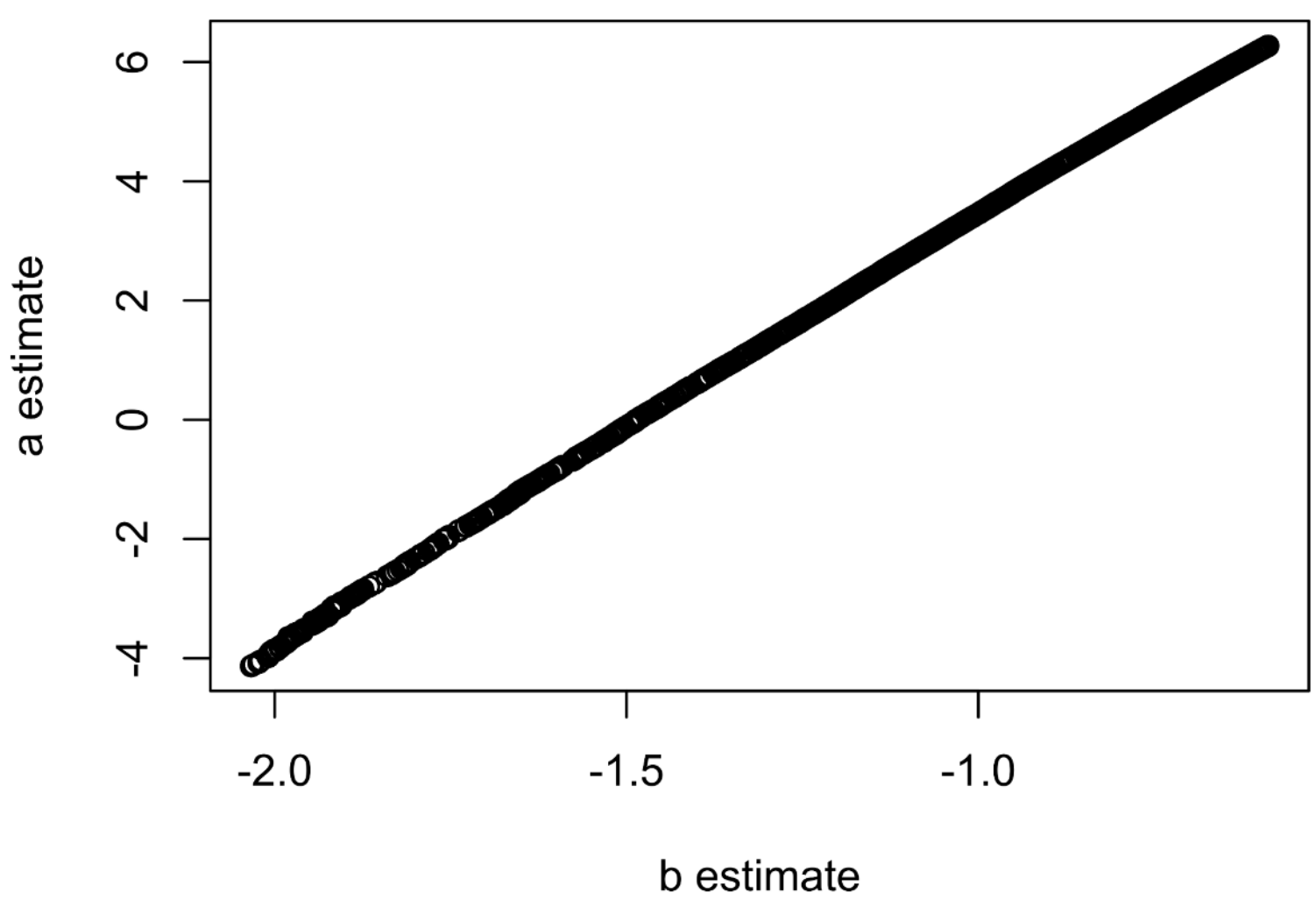

**Figure SM2.** Intercepts (a) vs slopes (b) of bNBS log-log-linear models. Note the near-perfect linear correlation between intercepts (a) and slopes (b). bNBS: backtransformed normalized biomass spectrum. N = 2000 simulations.

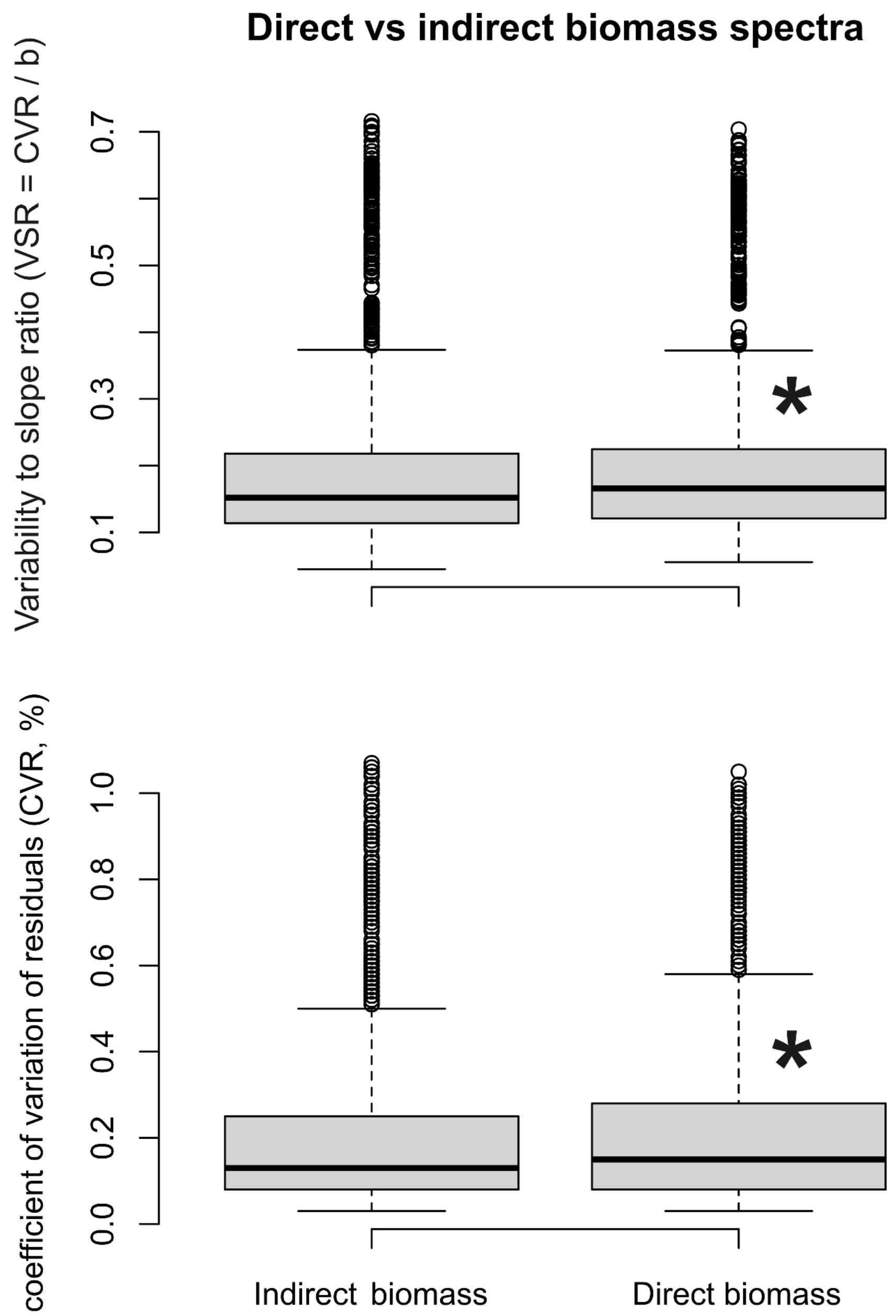

**Figure SM3.** Comparison of the variability represented in direct biomass spectra (direct binning of body mass data) *vs* indirect biomass spectra (biomass obtained through binned abundance, B = A * M). Note that the indirect method leads to lower represented variability (subestimation of variability) of bNBS linear models. bNBS: backtransformed normalized biomass spectrum. asterisks: differences in both plots are significant. Difference in medians: $p < 0.0001$ (permutation tests). Note the near-perfect linear correlation. n= 2000 simulations, low variability bNBS spectra, solution S1. Above: variability-to-slope-ratio, VSR (CVR / b), Below: coefficient of variation of residuals, (CVR, %).

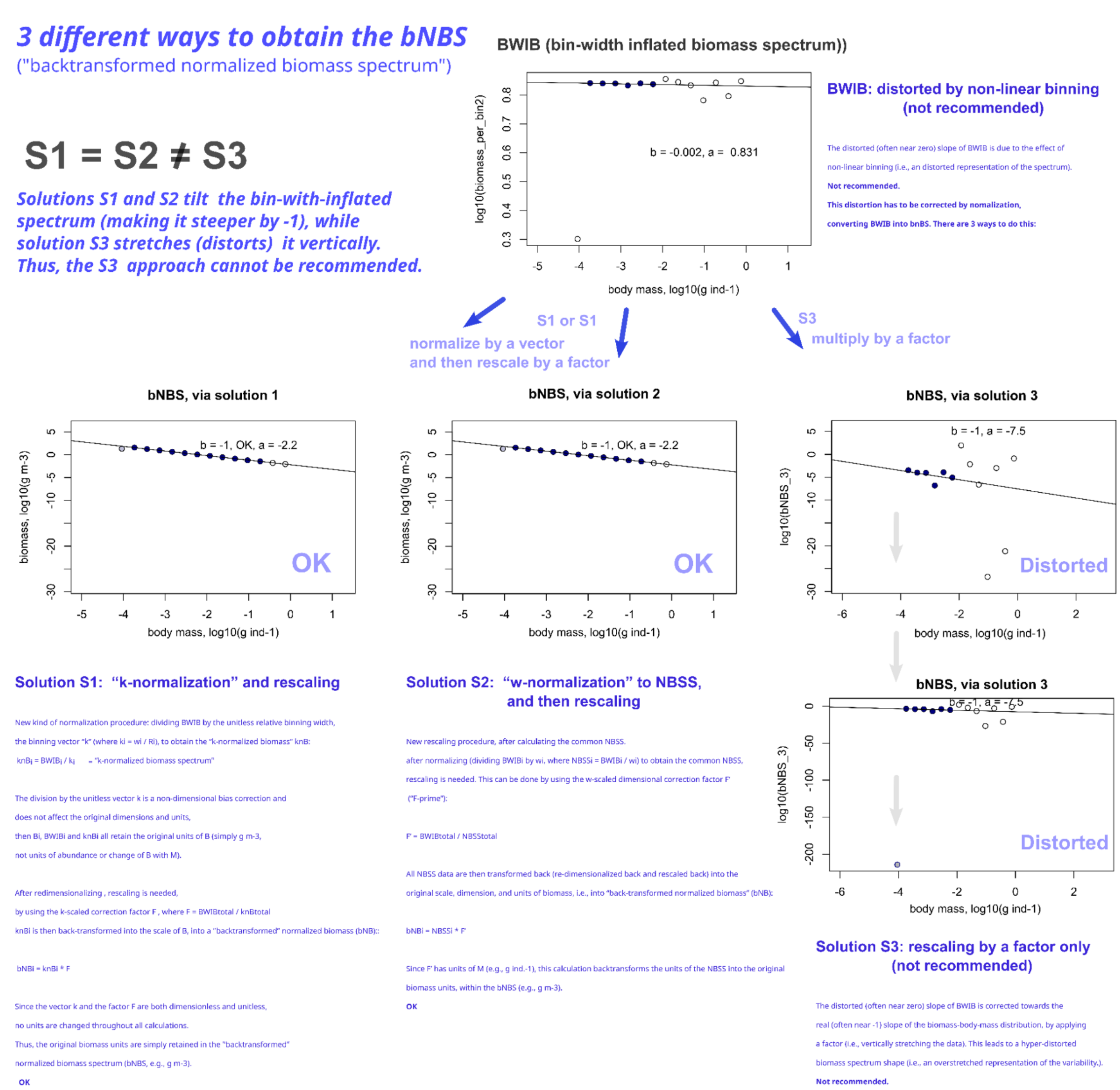

**Figure SM4a. Three ways (solutions S1, S2, and S3) to calculate the bNBS.** Examples of calculation with a low variability data set (perfect, low variability power law spectrum data, with beta = -1). Note that all 3 bNBS have the correct spectrum slope of b = - 1, and the correct bNBS units (g $m^{-3}$), but different shapes. bNBS: backtransformed normalized biomass spectra.

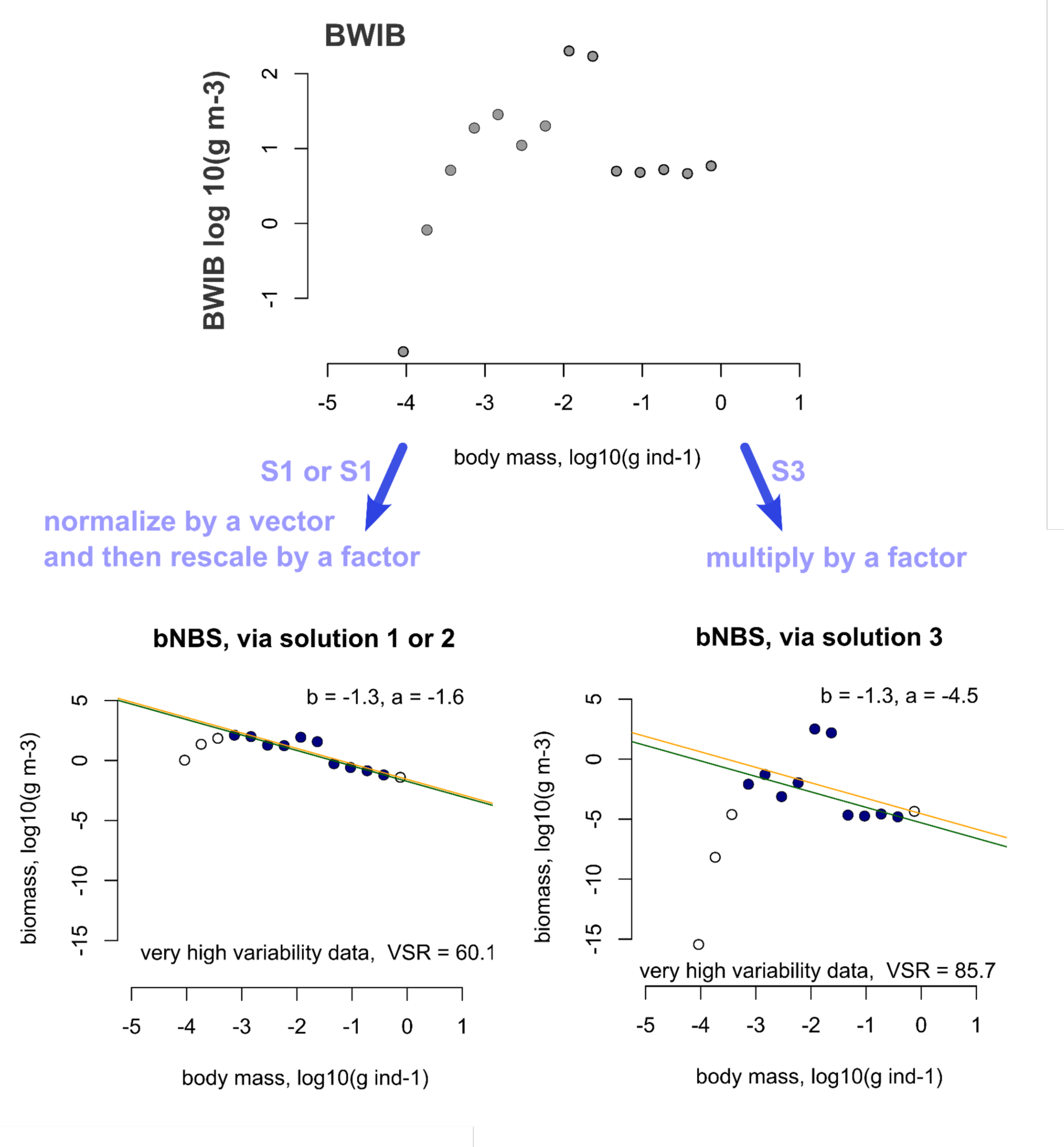

**Figure SM4b. Three ways (solutions S1, S2, and S3) to calculate the bNBS (high varability data).** Examples of calculation with a high variability data set (power law spectrum data with added eminent peaks). bNBS: backtransformed normalized biomass spectra. Orange: OLSR (ordinary least squares linear regression) linear models. Green line: robust regression linear models (parameters of the text in inlet refer to the robust regression model). Note that all bNBS have the same size spectrum slope "b", and the correct bNBS units (g $m^{-3}$), but different shapes.

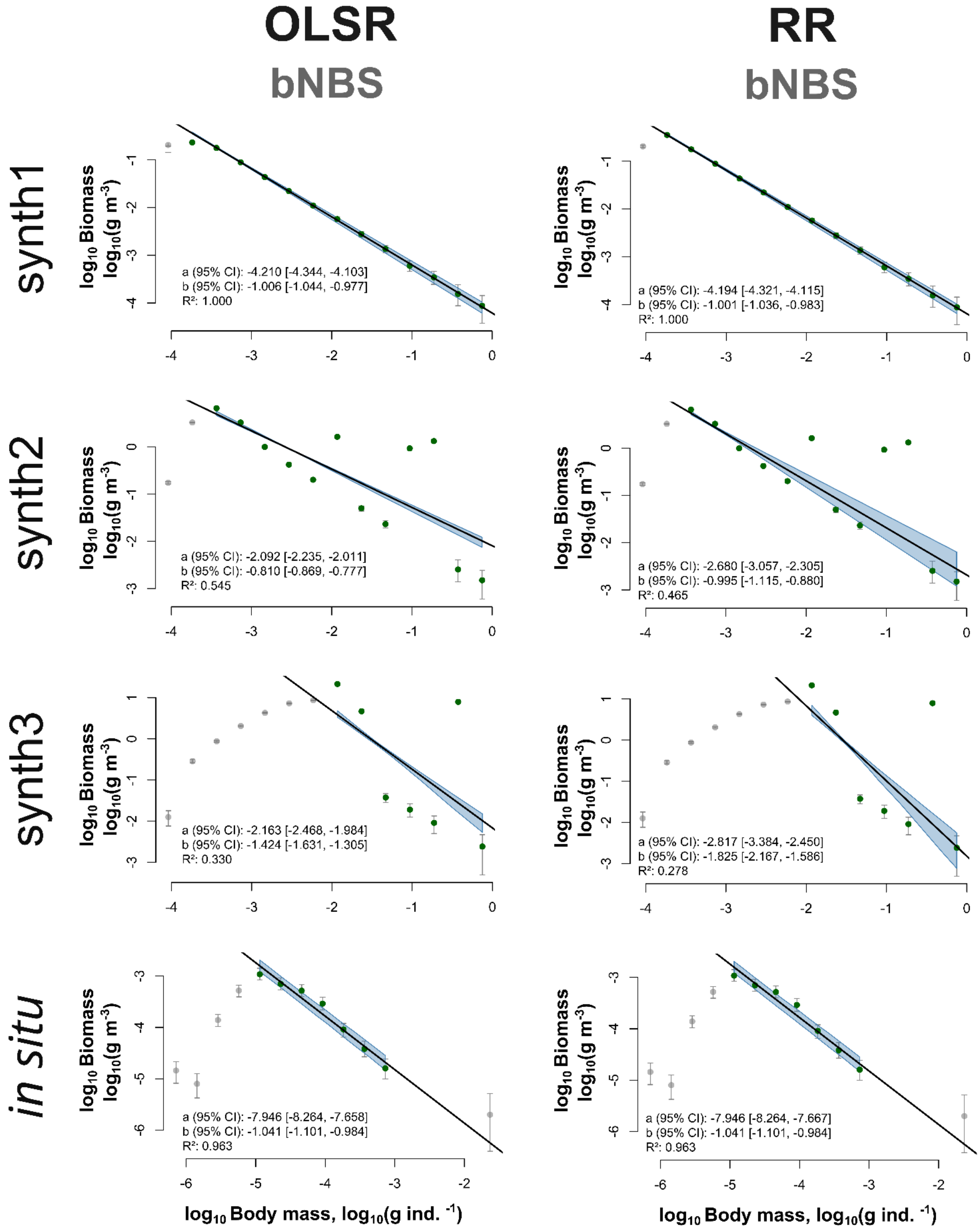

**Figure SM5a. OLSR vs RR, for bNBS.** Examples of bNBS (backtransformed normalized biomass spectra) that were fitted with OLSR (ordinary least squares linear regression) *vs* RR (robust regression). Example datasets used: syhth1: perfect power law distribution (near-zero variability); syhth2 : high-variability power law distribution data, with three peaks; syhth3: very high variability power law distribution data, with three sharp peaks; *in situ*: a tropical zooplankton sample, from Figueiredo et al. (2025). Vertical error bars: 95% bootstrap confidence intervals. Data for two-step boostrapped robust linear regression (blue 95% confidence envelopes) were selected from the maximum to the last non-zero bin.

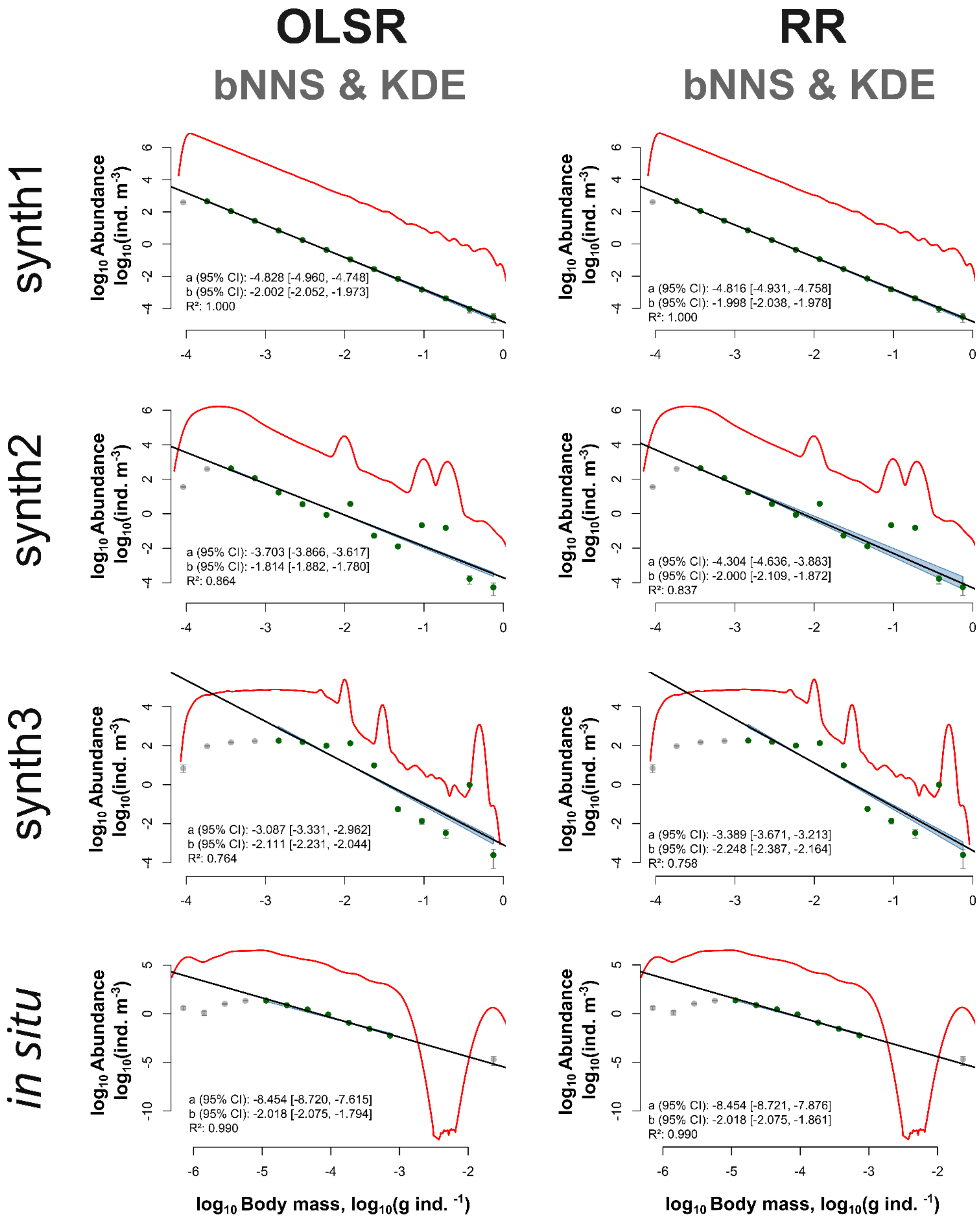

**Figure SM5b. OLSR vs RR, for bNNS.** Examples of bNNS (backtransformed normalized numbers spectra) that were fitted with OLSR (ordinary least squares linear regression) *vs* RR (robust regression). Red lines : KDE (kernel density estimation, bandwidth selection: Silverman's 'rule of thumb', Silverman, 1986); Example datasets used: syhth1: perfect power law distribution (near-zero variability); syhth2 : high-variability power law distribution data, with three peaks; syhth3: very high variability power law distribution data, with three sharp peaks; *in situ*: a tropical zooplankton sample, from Figueiredo et al. (2025). Vertical error bars: 95% bootstrap confidence intervals. Data for two-step boostrapped robust linear regression (blue 95% confidence envelopes) were selected from the maximum to the last non-zero bin.

**Table SMT1** Testing the NBSS (normalized biomass size spectrum) and NNSS (normalized numbers size spectrum)

| **Key Questions** | **Approaches for testing** | **Key Results and Conclusions** |
| --- | --- | --- |
| 1. Normalized Biomass = Biomass, Abundance, or Biomass flux? | 1. Compare normalized and non-normalized methods (linear binning, KDE), regarding size spectrum shape and slope. | 1 Normalized Biomass spectra (nonlinear binning and subsequent normalization) are similar in shape and slope to linear binning and KDE. Thus, NBSS represents biomass (not Abundance, or Biomass flux), and the currently used units must be corrected. |
| 2. Is the NBSS log-log-linear model slope OK? | 2. Calculate slope bias (input vs NBSS estimate, linear vs non-linear binning, KDE vs non-linear binning) | 2. NBSS slope is OK (negligible bias in all calculation methods tested) |
| 3. Variability well represented? Peaks and troughs well represented? Direct or indirect NBSS, which is best? | 3. Calculate bias in variability estimates (NNSS vs NBSS, linear vs non-linear methods, KDE vs non-linear binning, direct vs indirect binning). Compare KDE vs binning | 3. Best representations of the original variability: 1.) KDE with optimized bandwidth and 2.) direct binning with optimized bin widths. |
| 4. Are the calculations correct? Correct units and dimensions correctly transformed? | 4. Step-by-step verification (e.g. total biomass vs total BWIB) | 4. Total biomass sums are correct (total BWIB = Total Biomass), Calculations and transformations |

| | | of units are correct. |
|---|---|---|
| 5. How to modify (improve) the units and dimensions? Ways to obtain a back-transformation into the original units and dimensions? | 5. Step-by-step development, iterations, pruning, and testing | 5. Development of new size spectrum methods (for bNNS and bNBS). |
| 6. Find ways to improve the representation of variability in plots (get rid of linearizing artifacts and binning artifacts). | 6. Test three possible solutions for bNBS and bNNS calculations, and test KDE with different bandwidths. | 6. Solution S1 = Solution S2. Both solutions seem to work OK. Solution S3 produces a hyper-distorted BWIB (positive and negative slopes being over-inflated). KDE with SJ bandwidth selection best represents the shape (variability and slope) in the spectrum. |
| 7. Ways to allow quantitative comparisons of abundance and biomass across taxa, size classes, ecosystems, regions and time periods. | 7. Quantitative simulations, comparisons, and tests. | 7. Use a standardized binning vector, provide exact range and bin width for biomass values or abundance values at any chosen size or body mass (e.g., “y value at x = 0”, or “y value” at any other x value). Use a standardized factor to convert from NBSS to bNBS (e.g., a unity reference bin with $w_R$ = 1g) |